%% file: ANA-JETM-2024-06-PAPER.tex
\documentclass[cernpreprint, atlasdraft=false, UKenglish, texmf, orcidlogo]{atlasdoc}
\usepackage{atlaspackage}
\usepackage{atlasbiblatex}

\usepackage{atlasphysics}
\graphicspath{{logos/}{figures/}}

\usepackage{ANA-JETM-2024-06-PAPER-defs}

\AtlasTitle{Performance and efficiency of a transformer-based quark/gluon jet tagger in the ATLAS experiment}

\AtlasAbstract{%
A deep-learning approach based on the transformer architecture is developed to distinguish between jets originating from quarks and gluons.
The algorithm operates on jets with transverse momentum $p_{\text{T}} > 20$ and pseudorapidity $|\eta| < 4.5$ and takes as input several properties derived from the jet constituents, using information from the ATLAS detector's tracker and calorimeter.
The algorithm's performance is evaluated by analyzing dijet data events from proton--proton collisions at $\sqrt{s} = 13$ and $13.6$~TeV during Run~2 and Run~3 of the Large Hadron Collider.
Two methods are used to obtain distributions from quark- or gluon-initiated jets in data: a matrix method fully based on Monte Carlo simulation and a new approach named `jet topics' which has less dependence on the modelling of the physics process under study.
The quark and gluon identification efficiencies measured in data for the 50\% quark-identification-efficiency working point vary from the simulated ones for quark-initiated (gluon-initiated) jets by factors of 0.88--1.30 (0.61--1.05) with uncertainties of 10\%--70\% (10\%--95\%).
The uncertainties estimated with the jet topics method are smaller than those estimated with the matrix method, with up to 20\% less systematic uncertainty in some phase-space regions.
The advances in jet identification reported here provide a robust tool for precision Standard Model measurements and searches for new physics at the LHC.
}
\author{The ATLAS Collaboration}

\AtlasRefCode{JETM-2024-06}

\PreprintIdNumber{CERN-EP-2025-252}

\AtlasDate{\today}

\AtlasJournal{EPJC}
\hypersetup{pdftitle={ATLAS document},pdfauthor={The ATLAS Collaboration}}

\begin{document}

\maketitle

\tableofcontents

\section{Introduction}
%
%

Collisions at the Large Hadron Collider (LHC) \cite{Evans:2008zzb} produce high-energy quarks and gluons that hadronize into collimated sprays of particles called jets.
Theoretical studies have shown it is possible to label jets according to the type of parton that initiated their formation in an infrared- and collinear-safe manner~\cite{Banfi:2006hf,Caola:2023wpj}.
Moreover, it was observed that jet labelling is independent of the production process under consideration~\cite{Bright-Thonney:2018mxq}.
Once a labelling algorithm is chosen, jets labelled as originating from light quarks (quark jets) and the ones labelled as originating from gluons (gluon jets) can be studied, where gluon jets are seen to have a wider radiation pattern and be made of more particles with a lower average energy than quark jets~\cite{STDM-2017-16}.
These differences can be exploited to build algorithms ($q/g$ taggers) that infer the jet label and distinguish between the two jet types.
Given the process-independence of jet labelling, such taggers can be built to perform consistently across different physics analyses.
The precise definition of the jet labelling algorithm used in this analysis is given in \cref{sec:reco}.

Distinguishing between quark and gluon jets is crucial for a variety of applications.
For instance, many electroweak processes and new-physics scenarios predict signatures with quark jets~\cite{EXOT-2018-32,CMS-EXO-19-012,CMS-SUS-21-009,SUSY-2016-22}, while specific searches benefit from gluon jet identification~\cite{EXOT-2019-03,CMS-SUS-19-004}.
Measurements of Higgs boson properties exploit final states with quark jets produced by the vector-boson fusion process, and gluon jets arising from the gluon--gluon fusion process~\cite{HIGG-2023-11-v2,CMS-HIG-23-013}.
Furthermore, an accurate $q/g$ tagger can refine jet energy calibrations by using the different calorimeter responses to quark and gluon jets~\cite{JETM-2022-01}.

In the past, ATLAS has employed $q/g$ taggers based on variables associated with jets as a whole (high-level jet variables), such as the number of charged-particle tracks or the jet width~\cite{PERF-2013-02,ATL-PHYS-PUB-2017-009}.
More recent approaches have explored variables sensitive to energy and angular correlations between jet constituents \cite{Larkoski:2013eya} and even a convolutional neural-network approach using full jet images~\cite{ATL-PHYS-PUB-2017-017}.
ATLAS's most recent calibrated $q/g$ tagger was based on a Boosted Decision Tree architecture~\cite{JETM-2020-02}, taking high-level jet variables as input.
This paper presents a new $q/g$ tagger, trained using modern deep-learning techniques that directly process variables derived from jet constituents, building upon and extending an approach outlined in Ref.~\cite{ATL-PHYS-PUB-2023-032}.
The presented DeParT model, based on the ParT~\cite{part} and Transformer~\cite{vaswani2023attention} architectures, is applied across the full jet phase space reconstructed by ATLAS, consisting of jets with transverse momentum $\pt > 20~\gev$ and pseudorapidity $|\eta| < 4.5$.

The performance of the new $q/g$ tagger is evaluated using data from dijet events produced in proton--proton (\pp) collisions at $\sqrt{s} = 13~\tev$ and $13.6~\tev$ during \RunTwo and \RunThr of the LHC and recorded by ATLAS.
The tagger score distributions for quark and gluon jets are extracted from data using two methods: the jet topics method~\cite{Metodiev:2018ftz,Komiske:2022vxg,LeBlanc:2022bwd}, used to obtain the primary result of this paper, and the matrix method~\cite{ATL-PHYS-PUB-2017-009}, used in a previous ATLAS publication~\cite{JETM-2020-02} and taken as a reference.
Both methods exploit forward- and central-jet samples with different $q/g$ fractions to statistically demix them into samples of quark and gluon jets.
Jet topics is a novel technique that is expected to rely less on Monte Carlo (MC) simulations and provide a more precise evaluation of the $q/g$ tagger's performance.
The quark- and gluon-jet-tagging identification efficiencies are measured in data and compared with MC simulation predictions, at a fixed $q/g$ tagger working point.
Scale factors defined as the ratio of the measured identification efficiency to the expected identification efficiency are derived as a function of the jet \pt and \abseta.

This paper is organized as follows.
The ATLAS detector is described in Section~\ref{sec:detector}.
Section~\ref{sec:data} introduces the data sets used in the present publication.
Details of the reconstruction and identification of physical objects are provided in Section~\ref{sec:reco}.
The DeParT model is presented in Section~\ref{sec:tagger}.
Section~\ref{sec:extraction} describes the extraction of quark and gluon samples from data.
The scale factors are defined in Section~\ref{sec:sf_def} and Section~\ref{sec:uncerts} discusses their associated uncertainties.
The tagger's performance is summarized in Section~\ref{sec:results}, and Section~\ref{sec:conclusion} concludes the paper.


%
\FloatBarrier
\section{ATLAS detector}
\label{sec:detector}
%

\newcommand{\AtlasCoordFootnote}{%
ATLAS uses a right-handed coordinate system with its origin at the nominal interaction point (IP)
in the centre of the detector and the \(z\)-axis along the beam pipe.
The \(x\)-axis points from the IP to the centre of the LHC ring,
and the \(y\)-axis points upwards.
Polar coordinates \((r,\phi)\) are used in the transverse plane,
\(\phi\) being the azimuthal angle around the \(z\)-axis.
The pseudorapidity is defined in terms of the polar angle \(\theta\) as \(\eta = -\ln \tan(\theta/2)\) and is equal to the rapidity
$ y = (1/2) [(E + p_z)/(E - p_z)] $ in the relativistic limit.
Angular distance is measured in units of \(\Delta R \equiv \sqrt{(\Delta y)^{2} + (\Delta\phi)^{2}}\).}

The ATLAS detector~\cite{PERF-2007-01,GENR-2019-02} at the LHC covers nearly the entire solid angle around the collision point.\footnote{\AtlasCoordFootnote}
It consists of an inner tracking detector surrounded by a thin superconducting solenoid, electromagnetic and hadronic calorimeters,
and a muon spectrometer incorporating three large superconducting air-core toroidal magnets.

The inner-detector system is immersed in a \qty{2}{\tesla} axial magnetic field
and provides charged-particle tracking in the range \(|\eta| < 2.5\).
The high-granularity silicon pixel detector covers the interaction region and typically provides four measurements per track,
the first hit generally being in the insertable B-layer. %
It is followed by the silicon microstrip detector, which usually provides eight measurements per track.
These silicon detectors are complemented by the transition radiation tracker (TRT),
which enables radially extended track reconstruction up to \(|\eta| = 2.0\).
The TRT also provides electron identification information
based on the fraction of hits (typically 30 in total) above a higher energy-deposit threshold corresponding to transition radiation.

The calorimeter system covers the pseudorapidity range \(|\eta| < 4.9\).
Within the region \(|\eta|< 3.2\), electromagnetic calorimetry is provided by barrel and
endcap high-granularity lead/liquid-argon (LAr) calorimeters,
with an additional thin LAr presampler covering \(|\eta| < 1.8\)
to correct for energy loss in material upstream of the calorimeters.
Hadronic calorimetry is provided by the steel/scintillator-tile calorimeter,
segmented into three barrel structures within \(|\eta| < 1.7\), and two copper/LAr hadronic endcap calorimeters.
The solid angle coverage is completed with forward copper/LAr and tungsten/LAr calorimeter modules
optimized for electromagnetic and hadronic energy measurements, respectively.

The muon spectrometer comprises separate trigger and
high-precision tracking chambers measuring the deflection of muons in a magnetic field generated by the superconducting air-core toroidal magnets.
The field integral of the toroids ranges between \num{2.0} and \qty{6.0}{\tesla\metre} across most of the detector.
Three layers of precision chambers, each consisting of layers of monitored drift tubes, cover the region \(|\eta| < 2.7\).
These were complemented in the innermost layer of the endcap region by cathode-strip chambers in Run~2, which were replaced in Run~3 by layers of small-strip thin-gap chambers and Micromegas chambers, both providing precision tracking in the region \(1.3 < |\eta| < 2.7\).
The muon trigger system covers the range \(|\eta| < 2.4\) with resistive-plate chambers in the barrel region, thin-gap chambers in the endcap regions, and, in Run~3, the aforementioned small-strip thin-gap chambers and Micromegas chambers in the innermost layer of the endcap.

The luminosity is measured mainly by the LUCID-2 detector, which records Cherenkov light produced in the quartz windows of photomultipliers located close to the beampipe.

Events are selected by the first-level trigger system implemented in custom hardware,
followed by selections made by algorithms implemented in software in the high-level trigger~\cite{TRIG-2022-01}.
The first-level trigger accepts events from the \qty{40}{\MHz} bunch crossings at a rate close to \qty{100}{\kHz},
which the high-level trigger reduces in order to record complete events to disk at an average rate of about \qty{1.25}{\kHz} and  \qty{3}{\kHz} in \RunTwo and \RunThr, respectively.

Compared with its Run-2 configuration, the ATLAS detector benefits from several upgrades to maintain high performance at the higher \pileup levels in \RunThr.
The improvements include the new innermost layer of the muon spectrometer in the endcap region, which provides higher redundancy and a large reduction in fake muon triggers. The trigger system also benefits from new LAr calorimeter digital electronics with significantly increased granularity. These and other upgrades are described in more detail in Ref.~\cite{GENR-2019-02}.

A software suite~\cite{SOFT-2022-02} is used in data simulation, in the reconstruction
and analysis of real and simulated data, in detector operations, and in the trigger and data acquisition
systems of the experiment.


%
\FloatBarrier
\section{Data and simulation samples}
\label{sec:data}
%

This study uses data collected from proton--proton (\(pp\)) collisions by the ATLAS experiment during LHC \RunTwo in 2015--2018 at $\sqrt{s} = 13~\tev$ and LHC \RunThr in 2024 at $\sqrt{s} = 13.6~\tev$.
After imposing data quality requirements to ensure that the ATLAS detector was operating properly~\cite{DAPR-2018-01}, the integrated luminosities of the Run-2 and Run-3 data samples are $140.1 \pm 1.2~\ifb$~\cite{DAPR-2021-01} and $109.4 \pm 2.2~\ifb$~\cite{ATL-DAPR-PUB-2025-001}, respectively.

Monte Carlo simulations are used to model SM multijet production.
The generated events were passed through a \GEANT~\cite{Agostinelli:2002hh} simulation of the ATLAS detector and its response~\cite{SOFT-2010-01}.
The effect of multiple interactions in the same and neighbouring bunch crossings (\pileup) was modelled by overlaying~\cite{SIMU-2020-01} each simulated hard-scattering event  with a mix of inelastic \(pp\) scattering events generated by either  \EPOS[2.0.1.4]~\cite{Werner:2005jf} or \PYTHIA[8.308]~\cite{Bierlich:2022pfr}.
The \EPOS event generator used a set of tuned parameters called the \EPOS LHC tune~\cite{Pierog:2013ria}, while \PYTHIA used the A3 tune~\cite{ATL-PHYS-PUB-2016-017} and the \NNPDF[2.3lo]~\cite{Ball:2012cx} set of parton distribution functions (PDF).
Separate MC samples are used to model the different \pileup distributions observed in the 2015--2016, 2017, 2018 and 2024 data, and the simulated events are reweighted to reproduce the observed distribution of the average number of collisions per bunch crossing in each data-taking period.
All simulated events are reconstructed with the same software as the data and corrected so that the trigger, object reconstruction and object identification efficiencies, and the energy scales and resolutions, match the data.

\PYTHIA[8.235]~\cite{Sjostrand:2014zea} was used to produce the nominal simulation sample (referred to as the Pythia sample), which is based on dijet production matrix elements (ME) with leading-order (LO) QCD accuracy matched to a parton shower (PS) with leading-logarithm accuracy.
In the nominal settings, the renormalization and factorization scales are set to the geometric
mean of the squared transverse masses of the two outgoing particles in the matrix element.
The \NNPDF[2.3lo] PDF set was used in the ME generation, the parton shower, and the simulation of multiparton interactions.
The parton shower and multiparton interaction modelling uses parameter values from the A14 tune~\cite{ATL-PHYS-PUB-2014-021},
and the fragmentation and hadronization is based on the Lund string model~\cite{Andersson:1983ia,Sjostrand:1984ic}.
To populate the inclusive jet \pT spectrum efficiently, biased phase-space sampling was used in producing the Pythia sample; to compensate for this, the weighting of the simulated events was modified.

Alternative simulations that vary the modelling of dijet events are used to evaluate systematic uncertainties in the $q/g$ tagger's performance.
Uncertainties in perturbative calculations in the nominal simulation are propagated by using alternative weights for events in the Pythia sample.
The alternative weights cover variations of the scales at which the strong coupling constant is evaluated in the initial- and final-state showers, changes in the PDF choice, and variations in the non-singular part of the splitting functions~\cite{Mrenna:2016sih}.

The impact of next-to-leading-order (NLO) corrections in the ME is evaluated by using \POWHEGBOX[v2]~\cite{Nason:2004rx, Frixione:2007vw} to calculate the dijet production ME~\cite{Alioli:2010xa}, which is then interfaced with \PYTHIA[8.245].
This sample is referred to as Powheg+Pythia.
The \NNPDF[3.0nlo]~\cite{Ball:2014uwa} PDF set was used in the ME calculation, while the \NNPDF[2.3lo] set and the A14 tune were used to model the parton showers and multiparton interactions.

To evaluate uncertainties stemming from the choice of hadronization modelling, multijet events were produced with \SHERPA[2.2.5]~\cite{Bothmann:2019yzt} (referred to as Sherpa samples).
These samples use the  \CT[14nnlo] PDF set~\cite{Dulat:2015mca} and employ \(2\rightarrow2\) matrix elements at LO QCD accuracy in a showering algorithm based on Catani--Seymour dipole factorization with \pt ordering~\cite{Schumann:2007mg}.
One sample was produced with the \SHERPA AHADIC model for hadronization~\cite{Winter:2003tt} based on cluster fragmentation ideas and is referred to as the Sherpa Cluster sample, while another (referred to as the Sherpa String sample) was generated using the Lund string fragmentation model, which is also implemented in \PYTHIA[6]~\cite{Sjostrand:2006za}.

Multiple sets of multijet events were generated with \HERWIG~\cite{Bahr:2008pv,Bellm:2015jjp,Bellm:2017bvx} in order to assess how varying the PS algorithm changes the result.
For the 13~TeV simulation, events were produced with an angle-ordered shower evolution~\cite{Webber:1983if}, the \NNPDF[2.3lo] PDF set, and the Lund string fragmentation model of \PYTHIA[6]~\cite{Sjostrand:2006za}, where the Lund string implementation is used without tuning and does not include colour-reconnection effects (this sample is referred to as the Herwig String sample).
The 13.6~TeV simulation instead uses the \NNPDF[3.0nlo] PDF set and the cluster-based hadronization model with two shower configurations: one using the angle-ordered evolution (referred to as the Herwig Angular sample) and the other using the dipole algorithm \cite{Platzer:2009jq} (referred to as the Herwig Dipole sample).

The nominal and alternative MC samples are summarized in \cref{tab:mc_samples_calib}.

\begin{table}[!ht]
\centering
\caption{Nominal (first row) and alternative MC dijet samples with their configurations.
The table shows PDF sets, generator versions, QCD accuracy (Order), parton shower algorithms (PS alg.), hadronization models (Had.), and centre-of-mass energies used in a given sample.
}
\label{tab:mc_samples_calib}
\begin{tabular}{@{}lllllll@{}}
\toprule
Name & PDF set & Generator & Order & PS alg. & Had. & $\sqrt{s}$ [TeV] \\ \midrule
Pythia & \NNPDF[2.3lo] & \Pythia[8.235] & LO & $\pT$-ord. & String & 13 / 13.6 \\
Powheg+Pythia & \NNPDF[3.0nlo] & \POWPY[8] & NLO & $\pT$-ord. & String & 13 / 13.6 \\
Sherpa Cluster & \CT[14nnlo] & \Sherpa[2.2.5] & LO &  $\pT$-ord. & Cluster & 13 / 13.6 \\
Sherpa String & \CT[14nnlo] & \Sherpa[2.2.5] & LO & $\pT$-ord. & String & 13 / 13.6 \\
Herwig String & \NNPDF[2.3lo] & \Herwig[7] & LO & Angle-ord. & String & 13  \\
Herwig Angular & \NNPDF[3.0nlo] & \Herwig[7] & LO & Angle-ord. & Cluster & 13.6 \\
Herwig Dipole & \NNPDF[3.0nlo] & \Herwig[7] & LO & Dipole & Cluster & 13.6 \\
\bottomrule
\end{tabular}
\end{table}



%
\FloatBarrier
\section{Object reconstruction and event selection}
\label{sec:reco}
%

Jets are reconstructed from particle-flow objects (PFOs)~\cite{PERF-2015-09} by the \antikt algorithm~\cite{Cacciari:2008gp} with a radius parameter of $R = 0.4$ as implemented in FastJet~\cite{Fastjet}.
The PFOs are a collection of reconstructed objects formed by the particle-flow algorithm~\cite{PERF-2015-09}, which systematically combines charged-particle tracks and calorimeter objects called topological clusters (topoclusters)~\cite{PERF-2014-07}.
Specifically, PFO momenta are obtained by replacing the energy deposited by charged particles in a region of the calorimeter with the momenta of tracks that are matched to that specific calorimeter region.
When some energy remains in the calorimeter region or when no matching track is found, neutral PFOs are created from the corresponding topoclusters.
An overall jet-energy calibration~\cite{JETM-2018-05} accounts for residual detector effects as well as contributions from \pileup.

In addition to PFOs, TopoTowers are employed as auxiliary jet variables.
Calorimeter towers \cite{PERF-2011-03} are static grid elements with a size of $\Delta \eta \times \Delta \phi = 0.1 \times 0.1$, constructed from calorimeter cells.
TopoTowers are calorimeter towers for which the only contributing cells are those associated with an energy cluster.
While PFOs are used throughout the jet reconstruction and calibration chain, TopoTowers are ghost-associated~\cite{Cacciari:2007fd} to jets and serve as supplementary input to the $q/g$ tagger, providing higher-granularity calorimeter information in the forward region where tracking detector coverage is limited or absent.
For jets fully outside the tracking acceptance ($|\eta| > 2.9$), there are on average 21 TopoTowers per PFO.
TopoTowers are included only for $|\eta| > 2.0$ and are particularly impactful for $|\eta| > 3.4$.
In this region, significant ambient energy is present and the forward calorimeter geometry is not aligned with an $\eta\times\phi$ representation; TopoTowers can therefore split individual topoclusters covering a large part of the sub-detector.

In each event of the MC simulation, a second set of $R = 0.4$ jets (called \enquote{truth} jets) is built with the \antikt algorithm directly from the generated stable final-state particles\footnote{Particles are considered stable if their decay length is greater than 1~cm.} (excluding muons, neutrinos, and particles from pile-up interactions) before they interact with the detector.
Information about the parton that initiated the jet is retrieved~\cite{PERF-2013-02} from the parton shower record by identifying the highest-energy parton ghost-associated~\cite{Cacciari:2007fd} with the truth jet.
Truth jets are geometrically matched to the reconstructed ones by requiring their angular separation to be $\Delta R < 0.3$.
No additional discriminating variables are used to differentiate heavy-flavour quark jets from light-flavour quark jets~\cite{CDF:2008ixu,STDM-2011-48}.
When either a $b$-quark or $c$-quark is the highest-energy parton, the jet receives a quark jet label.
Heavy-flavour jets comprise less than 2\% of the total dataset across all considered \pt bins.
Reconstructed jets remain unlabelled if no truth parton with $\pt > 1$~\GeV\ is found among the partons ghost-associated with the truth jet, or if they arise from \pileup.
Unlabelled jets comprise less than 1\% of the dataset considered, and they are thus ignored~\cite{STDM-2011-40}.

Events used in the analysis were selected with a set of single-jet triggers~\cite{TRIG-2019-04,TRIG-2022-01}.
The trigger strategy employs two complementary categories: central-jet triggers with detector pseudorapidity acceptance $|\eta| < 3.2$ and forward-jet triggers with $|\eta| > 3.2$.
The central-jet triggers utilize \pt thresholds ranging from $35$~\GeV to $525$~\GeV, while the forward-jet triggers employ thresholds from $35$~\GeV to $350$~\GeV.
Additionally, a single-jet trigger with no \pt threshold is included within the whole $|\eta|$ detector acceptance to allow jet \pt coverage down to 20~\GeV in both the central and forward regions.
Single-jet triggers with low \pt thresholds are prescaled to reduce the high trigger rates and record only a fraction of the events for online processing and data storage at an acceptable level.
Only the highest-\pt-threshold triggers (both central and forward) are unprescaled, so as to record each observed event containing high-\pt jets.
A leading-jet trigger strategy~\cite{Lendermann_2009} is employed, where the trigger is accepted only if the leading (i.e.\ highest-\pt) jet fell within a predefined \pT range, achieving a trigger efficiency of at least 99.5\%.
The \pT thresholds differ between central- and forward-jet triggers, so the leading jet is classified according to its \abseta to determine the appropriate \pT range for trigger selection.
To obtain a smooth \pt spectrum at reconstruction level, events are weighted according to the trigger prescale factors.

The offline selection requires events to contain at least two jets with $20~\GeV < \pt < 2$~\TeV and $|\eta| < 4.5$, ensuring complete containment within the calorimeter acceptance.
The jet vertex tagger (JVT) \cite{PERF-2014-03} is used to remove \pileup jets with $\pt < 60$~\GeV and $\abseta < 2.5$ that are identified as not associated with the event's primary vertex.
Similarly, pile-up jets in the forward region are suppressed with a `forward JVT' (fJVT) \cite{PERF-2016-06} algorithm, which exploits jet shapes and topological jet correlations in pile-up interactions and is applied to all jets with $\pt < 60$~\GeV and $\abseta > 2.5$.
Simulated events are corrected for differences between data and MC simulation seen in JVT and fJVT efficiencies \cite{PERF-2014-03,PERF-2016-06}.
The two highest-\pt jets ($j_1$ and $j_2$) must have azimuthal separation $\Delta\phi(j_1,j_2) > 2.8$ and if a third jet exists in the event its \pt is required to be smaller than 0.05 times the sum of the \pt values of $j_1$ and $j_2$, to better select events with a dijet topology.
The performance of the $q/g$ tagger is evaluated in bins of jet $\pT$ and $\eta$, with at least one jet required to satisfy the conditions on $\pT$ and $\eta$ to belong to a certain bin.


%
\FloatBarrier
\section{Tagger definition}
\label{sec:tagger}
%

The $q/g$ tagger considered in this paper is based on the DeParT model~\cite{ATL-PHYS-PUB-2023-032,part}.
The tagger is trained on the two highest-\pt jets in events generated in the nominal \Pythia simulation with Run-2 conditions.
It is preferred that the tagger is not reliant on the \pt and $\eta$ of the jet for the $q/g$ classification,
allowing physics analyses to use either the full jet spectrum or a subset of it.
Due to the limited size of the simulated event samples, the dijet events do not cover the entire phase space spanning 20--2000~GeV in \pt and $|\eta| < 4.5$.
To overcome this problem, an ensemble of three subsamples covering different $ \pt $ and $ \eta $ ranges is used for training, and each subsample is reweighted to produce a uniform two-dimensional distribution in $ \pt \times \eta $ space.
This ensures a flat distribution of jet \pt and $\eta$ in the two dimensions, and the flattening procedure for each subsample is performed separately for quark and gluon jets to avoid a tagger bias towards one of the jet types.
Each subsample is normalized such that the sum of weights is equal to the number of events in that subsample, and the total number of jets in the training ensemble is 102\,M.
The number of jets in each subsample is shown along with the corresponding \pt and $|\eta|$ ranges in \cref{tab:datasets}.
A statistically independent validation sample containing 10.2\,M jets is constructed in the same way as the training sample and is used for model evaluation during training.
Overfitting is prevented by evaluating the loss on this sample after each training epoch.
The validation loss is monitored throughout training and does not diverge from the trend observed for the training loss, confirming the absence of overfitting.

\begin{table}[htb!]
\centering
\caption{The \pt and $\eta$ ranges of the three subsamples used to train the tagger and their relative and absolute sizes.}
\label{tab:datasets}
\begin{tabular}{lcccccc}
\hline
Index & \pT range & $\eta$ range & Total size & Relative size \\
\hline
1.   & $20 < \pt < 160$ GeV & $|\eta| < 4.5$ & 50\,M & 49\% \\
2.   & $160 < \pt < 1300$ GeV & $|\eta| < 2.1$ & 50\,M & 49\%  \\
3.   &  $1300 < \pt < 2000$ GeV~~ & $|\eta| < 1.2$ & ~~~~\,2.1\,M & ~~2\% \\
\hline
\end{tabular}
\end{table}

The DeParT network takes as input several variables constructed from the jet constituents, namely the PFOs and TopoTowers.
Such variables can be separated into constituent variables, where the individual constituent momenta are taken separately, and constituent interaction variables, where pairs of constituent momenta are considered.

\begin{table}[!htb]
\centering
\caption{Constituent variables used as input to DeParT. }
\label{tab:const_variables}
\begin{tabular}{l}
\toprule
Constituent Variables \\
\midrule
$\log(p_\mathrm{T}/p_\mathrm{T}^\mathrm{jet})$ \\
$\log(E/p_\mathrm{T}^\mathrm{jet})$ \\
$\Delta \eta = \eta - \eta^\mathrm{jet}$ \\
$\Delta \phi = \phi - \phi^\mathrm{jet}$ \\
$\Delta R = \sqrt{(\Delta \eta)^2 + (\Delta \phi)^2}$ \\
isCharged \\
isTopo \\
\bottomrule
\end{tabular}
\end{table}

Constituent variables include the logarithms of the ratios of the constituent's $\pT$ or energy $E$ to the jet's \pt; the pseudorapidity difference $\Delta\eta$ between the constituent and the jet axis; the azimuthal angle $\Delta\phi$ between the constituent and the jet axis; and the angular separation $\Delta R$ between the constituent and the jet axis.
The logarithms of the constituent's \pt and energy are not included, unlike in Refs.~\cite{part,ATL-PHYS-PUB-2023-032}, since it was found that they do not improve the tagger's performance.
A binary variable called `isCharged' is used to distinguish between charged and neutral constituents: its value is one for PFOs with an associated track and zero for neutral PFOs and TopoTowers.
An overlap between PFOs and TopoTowers is resolved by using the `isTopo' constituent variable, which allows the model to learn the relationship between PFOs and TopoTowers through the attention mechanism during training.
The value of `isTopo' is zero for PFOs and one for TopoTowers.
The complete list of constituent variables is shown in \cref{tab:const_variables}.%

\begin{table}[!htb]
\centering
\caption{List of constituent interaction variables used as input to DeParT.
Indices $a$ and $b$ indicate the two constituents for which the interaction variables are calculated.
}
\label{tab:const_interaction_variables}
\begin{tabular}{l}
\toprule
Constituent Interaction Variables \\
\midrule
$\log \Delta^{ab} = \log{\sqrt{(\eta^a - \eta^b)^2 + (\phi^a - \phi^b)^2}}$ \\
$\log m^{2, ab} = \log{\left[(p^{a} + p^{b})^2/(p_\mathrm{T}^{\mathrm{jet}})^2\right]}$ \\
$\log k_\mathrm{T}^{ab} = \log{\left[\min(p_\mathrm{T}^a/p_\mathrm{T}^{\mathrm{jet}}, p_\mathrm{T}^b/p_\mathrm{T}^{\mathrm{jet}}) \Delta^{ab}\right]}$ \\
$z^{ab} = \min{(p_\mathrm{T}^a, p_\mathrm{T}^b)}/(p_\mathrm{T}^a + p_\mathrm{T}^b)$ \\
\bottomrule
\end{tabular}
\end{table}

The constituent interaction variables are calculated for each pair ($a$, $b$) of constituents in a given jet.
The variable $\Delta^{ab}$ represents the angular separation between two jet constituents.
The invariant mass of the two constituents, $m^{2, ab}$, is calculated from their four-momenta and scaled by the total jet \pt.
The $k_\mathrm{T}^{ab}$ variable characterizes the transverse momentum scale of the splitting, defined as the product of the softer constituent's momentum fraction and the angular separation $\Delta^{ab}$ between the two constituents.
Unlike in Refs.~\cite{ATL-PHYS-PUB-2023-032,part}, the constituents' \pt fractions rather than their absolute \pt are used in the $k_\mathrm{T}^{ab}$ variable; this reduces the overall \pt dependence of the tagger.
The last variable, $z^{ab}$, quantifies the momentum sharing between constituents, representing the fraction of their combined transverse momentum carried by the softer constituent.
The complete list of constituent interaction variables is shown in \cref{tab:const_interaction_variables}.%

Before being fed into the tagger, the distributions of all the variables were normalized to have a mean of zero and a standard deviation of one, ensuring that features at different natural scales contribute equally during training.

Due to GPU memory constraints and high memory usage by the transformers, only the 60 highest-\pt constituents of each jet are considered.
This constraint affects only about 5.7\% of the jets in a testing dataset with the physical $p_{\text{T}}$ and $\eta$ spectrum and 8.6\% in the training dataset,\footnote{The difference between these fractions in the two datasets is attributed to the reweighting procedure applied to the training dataset.} and it was found not to have any impact on the tagger's performance.

The DeParT model is trained for 10 epochs with a batch size of 1024.
This number of epochs proved sufficient given the large size (102\,M jets) of the dataset, with the model typically reaching minimum validation loss before completing all epochs.
Training for additional epochs did not improve performance on the validation dataset.
The Adam optimizer \cite{adam} is employed with a learning rate increasing linearly from zero to $10^{-4}$ for the first 20\,k training steps, followed by cosine decay to 1\% of the maximum learning rate over the subsequent 10 epochs.
The architectural hyperparameters of the DeParT model are the same as those used in Ref.~\cite{ATL-PHYS-PUB-2023-032}, and the total number of trainable parameters is 2.6~million.
The training dataset size of 102\,M jets provides sufficient statistical power to effectively constrain the 2.6~million model parameters and avoid overfitting.

\begin{figure}[!htb]
\subfloat[]{
\includegraphics[width=0.49\linewidth]{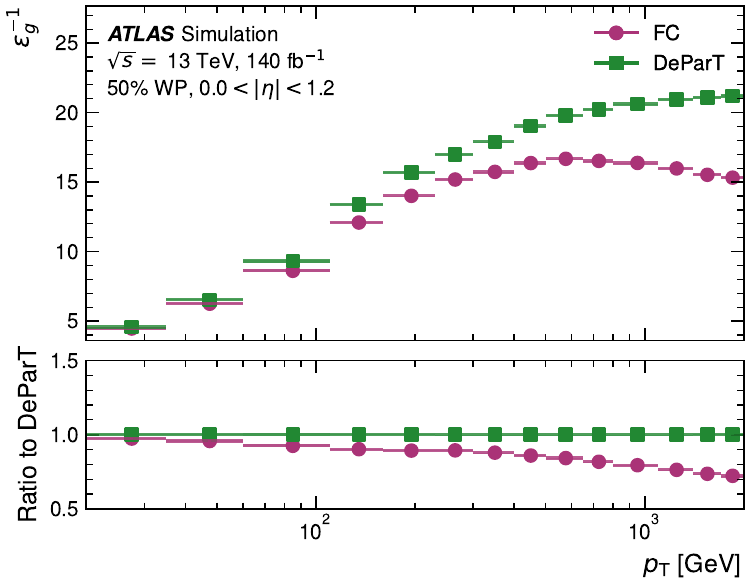}
\label{fig:fc_vs_depart_a}
}\subfloat[]{
\includegraphics[width=0.49\linewidth]{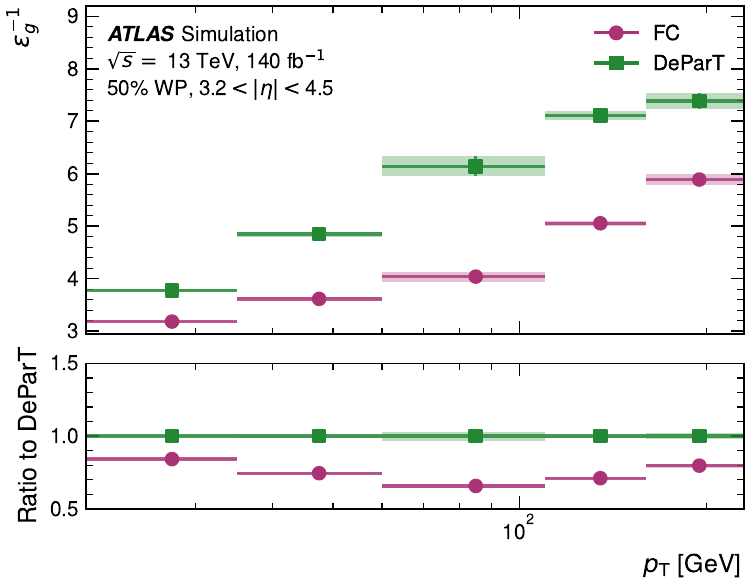}
\label{fig:fc_vs_depart_b}
}
\caption{The gluon jet rejection at 50\% quark jet efficiency as a function of \pt for jets with
\protect\subref{fig:fc_vs_depart_a}~$|\eta| = 0.0{-}1.2$ and
\protect\subref{fig:fc_vs_depart_b}~$|\eta| = 3.2{-}4.5$ for the FC and DeParT models.
The bottom panels show the FC-to-DeParT ratios.
The error bars represent the statistical uncertainties.
}
\label{fig:fc_vs_depart}
\end{figure}

The tagger's performance evaluated on the nominal Pythia sample with the physical \pt and $\eta$ spectrum is shown in \cref{fig:fc_vs_depart}.
The efficiency of the quark (gluon) jet tagging, denoted by $\varepsilon_q$ ($\varepsilon_g$), is the ratio of the number of correctly (incorrectly) classified quark (gluon) jets to the total number of quark (gluon) jets in the used sample.
The symbol $\varepsilon_g^{-1}$, the inverse of $\varepsilon_g$, stands for the rejection of gluon jets.
The metric used for comparison is $\varepsilon_g^{-1}$ at a fixed $\varepsilon_q$ value of 0.5, i.e.\ the 50\% working point (WP),
as a function of \pt for two $|\eta|$ bins.
The score threshold defining the 50\% WP is determined from the Pythia sample in each \pt and $|\eta|$ bin separately.
The performance of the DeParT model is compared with that of a fully connected (FC) neural network with residual connections~\cite{srivastava2015highway}, which takes as input the jet variables defined in Ref.~\cite{ATL-PHYS-PUB-2023-032}.
The FC network serves as a baseline because of its competitive performance among previously developed taggers.
The classification performance is seen to improve with increasing \pt for both models.
This is expected because at higher jet \pt more jet constituents are available for feature extraction.
DeParT performs better than the FC model in most \pt bins and $|\eta|$ regions.
In the central region, shown in \cref{fig:fc_vs_depart_a}, the additional complexity of the DeParT model is seen to gradually outperform the FC network with increasing jet \pt, taking advantage of the kinematic regime where the jets contain more constituents.
The  performance difference becomes more pronounced at higher \pt values in the central region.
The TopoTowers incorporated in the DeParT architecture provide advantages in the forward region, with $\varepsilon_g^{-1}$ exceeding the FC baseline value even at low \pt, as shown in \cref{fig:fc_vs_depart_b}.

\begin{figure}[!htb]
\subfloat[]{
\includegraphics[width=0.49\linewidth]{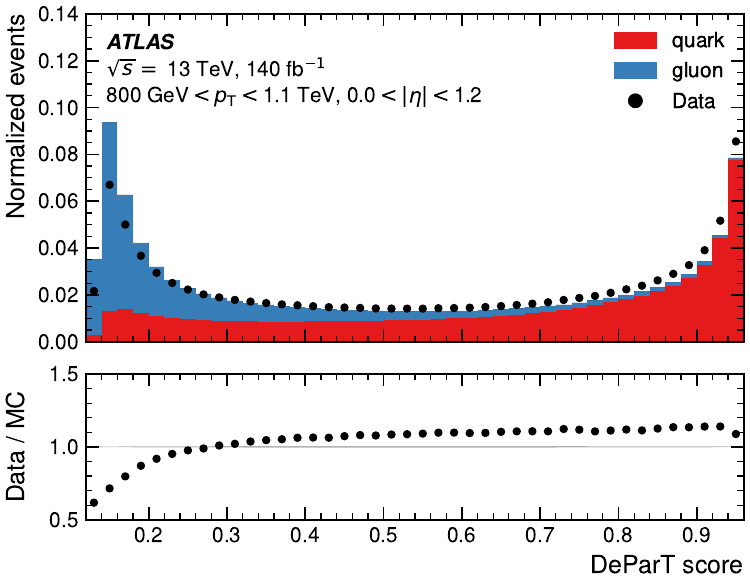}
\label{fig:score_mc_data_run2_ctrl}
}\subfloat[]{
\includegraphics[width=0.49\linewidth]{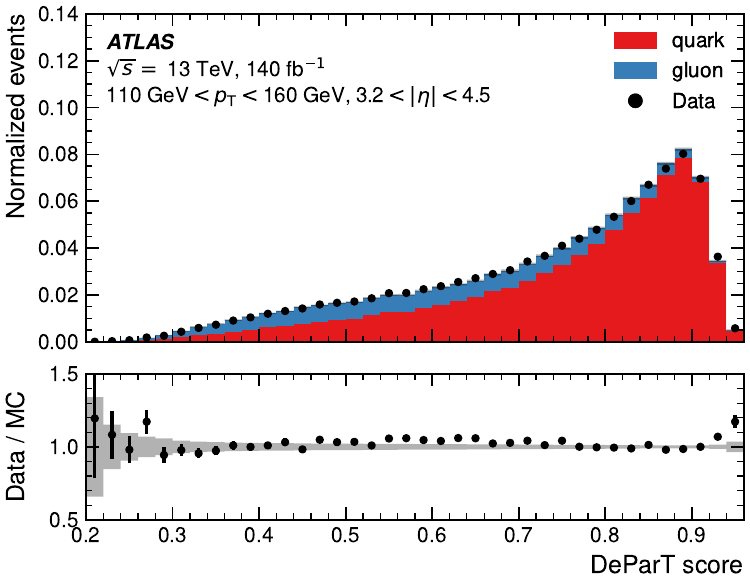}
\label{fig:score_mc_data_run2_fwd}
}\\
\subfloat[]{
\includegraphics[width=0.49\linewidth]{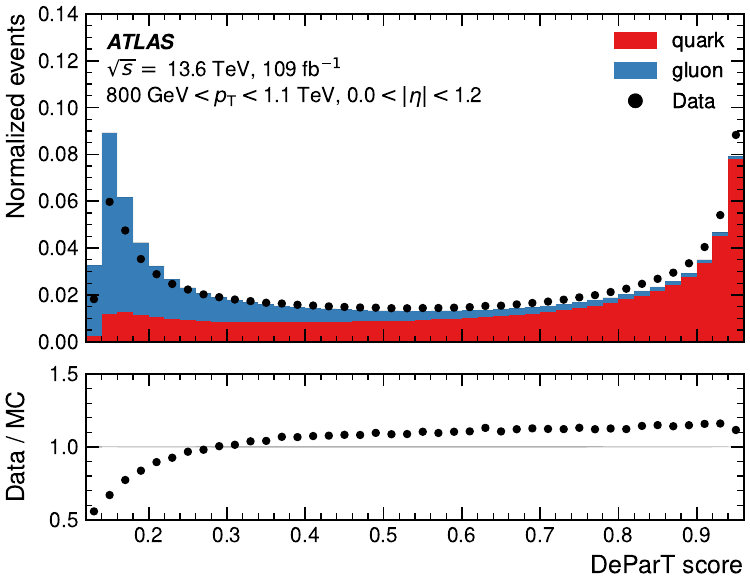}
\label{fig:score_mc_data_run3_ctrl}
}\subfloat[]{
\includegraphics[width=0.49\linewidth]{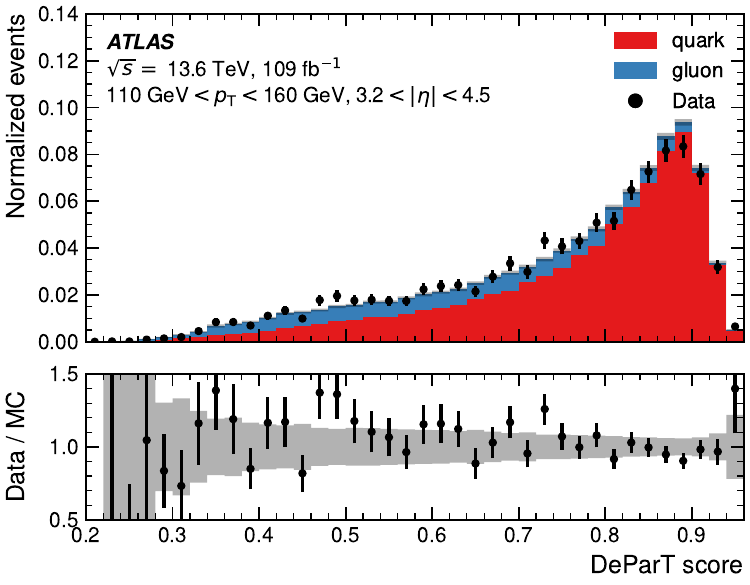}
\label{fig:score_mc_data_run3_fwd}
}
\caption{The DeParT score distributions for quark and gluon jets in MC events compared with
\protect\subref{fig:score_mc_data_run2_ctrl}~Run-2 and \protect\subref{fig:score_mc_data_run3_ctrl}~Run-3 jets with $  0.0 < |\eta| < 1.2$ and $ 800 < \pt < 1100\ \gev$, and
\protect\subref{fig:score_mc_data_run2_fwd}~Run-2 and \protect\subref{fig:score_mc_data_run3_fwd}~Run-3 jets with $3.2 < |\eta| < 4.5$ and $ 110 < \pt < 160\ \gev$.
The error bars represent the statistical uncertainties of data and the shaded bands represent the statistical uncertainties of MC events.
The bottom panels show the data-to-MC ratios.
}
\label{fig:score_mc_data}
\end{figure}

The DeParT score distributions for the jets selected from the Pythia sample and the measured data are presented in \cref{fig:score_mc_data}, for different jet kinematic ranges.
The pronounced peaks observed around 0.95 (0.15) for MC quark (gluon) jets in \cref{fig:score_mc_data_run2_ctrl,fig:score_mc_data_run3_ctrl} demonstrate the discrimination power of the tagger.
The tagger scores for quark and gluon jets do not peak at the extreme values of 0 and 1, respectively, due to two contributing factors: the application of physical weights, in contrast to the flattening weights employed during the training phase, and the kinematic selection of jets within a specific \pt and \abseta range.
The peak in tagger score closer to 0 is seen to be significantly smaller in data than in MC simulation, reaching only 60\% of the peak height in the simulation.
In the forward pseudorapidity region (\cref{fig:score_mc_data_run2_fwd,fig:score_mc_data_run3_fwd}), the distribution is dominated by a large peak near unity corresponding to quark jets.
In contrast to the central region, the gluon jet component does not exhibit a prominent peak at low scores but instead shows a more uniform distribution across the tagger output range.
Nevertheless, the separation between the quark and gluon score distributions remains sufficient to enable discrimination in this forward kinematic regime.
Agreement between experimental data and MC simulation is much better in the forward region than in the central detector region.
The observed differences between Run-2 and Run-3 datasets are not considered significant, with variations in Run-3 data attributable to lower statistical precision rather than systematic effects.
The larger statistical uncertainties in the Run-3 data are due to the dataset being smaller than in \RunTwo and the use of different trigger prescale factors.


%
\FloatBarrier
\section{Extraction of quark and gluon distributions in data}
\label{sec:extraction}
%

Quark-only and gluon-only jet samples are necessary to measure the performance of a $q/g$ tagger.
In the context of the definition in \cref{sec:reco}, this can be accomplished in the MC simulation by using the underlying truth information associated with the jet, while for data this is of course not possible.
Two extraction methods are used to obtain quark and gluon tagger score distributions in data.
Both methods exploit the different $q/g$ fractions in two samples, either quark- or gluon-enriched, and demix them into quark and gluon distributions.

Quark- and gluon-enriched samples are defined from the two highest-\pt jets in each event: the jet with the higher \abseta value is selected to populate the quark-enriched sample and the one with lower \abseta is assigned to a gluon-enriched sample.
This classification strategy leverages the fact that in the high proton-momentum-fraction range, the PDF has a higher probability of including valence quarks.
Consequently, the ensemble of jets which are more forward (higher \abseta) have a higher probability of being quark jets, whereas the ensemble of jets which are more central (lower \abseta) have a higher probability of being gluon jets~\cite{STDM-2014-17}.
The fraction of dijet events where the two highest-\pt jets consist of one quark jet and one gluon jet ranges approximately from 36\% to 68\% depending on the jet \pt and $\eta$.
The quark-enriched (gluon-enriched) sample is referred to as the forward (central) sample in the following.

Jet fragmentation is expected to be $\eta$-independent at the particle level.
Consequently, the distributions of the tagger score $x$ for quark jets, $p_q(x)$, and gluon jets, $p_g(x)$, are assumed identical in the central and forward samples.
This allows the central distribution $p_\text{C}(x)$ and the forward distribution $p_\text{F}(x)$ as
\begin{align}
p_\text{C}(x) = f_\text{C}^q p_{q}(x)+(1 - f_\text{C}^q) p_{g}(x)\label{eq:forward_central_1}\ , \\
p_\text{F}(x) = f_\text{F}^q p_{q}(x)+(1 - f_\text{F}^q) p_{g}(x)\label{eq:forward_central_2}\ ,
\end{align}
where $f_\text{C}^q$ and $f_\text{F}^q$ are the fractions of quark jets in the central and forward samples, respectively.

\subsection{Central-to-forward sample reweighting}
\label{sec:reweighting}
As mentioned in the previous section, the score distributions are expected to be identical at particle level, but detector-level measurements can depend on $\eta$ because of changes in the type of detector technology or the amount of detector material traversed.
These detector effects cause the $p_{q}(x)$ and $p_{g}(x)$ distributions to each have different shapes in the central and forward samples, resulting in non-closure of the extraction methods described below.
To ensure that the shapes of the $p_{q}(x)$ and $p_{g}(x)$ distributions are independent of whether they are evaluated in the central or forward sample, a reweighting procedure is applied in bins of jet \pt and $\eta$ as described in Ref.~\cite{JETM-2020-02}.
Various possibilities exist for implementing this reweighting: the central distributions can be adjusted to match the forward ones or vice versa, and the weights can be derived from either the quark or gluon distributions.
As a nominal choice, weights are derived from MC simulations to match the central quark distribution to the forward quark distribution, with these MC-derived weights subsequently applied identically to both data and MC central distributions.
The alternative reweightings are used to estimate the uncertainties associated with this choice.
The values of the resulting weights range approximately from 0.5 to 2.7 for well-populated bins of the score distributions with a statistical uncertainty below 20\%.
The reweighting procedure is performed before applying either of the two demixing methods described below.

\subsection{Matrix method}
The first method used to extract quark and gluon distributions from data is the matrix method~\cite{ATL-PHYS-PUB-2017-009,JETM-2020-02}, which assumes the fractions $f_\text{C}^q$ and $f_\text{F}^q$ from \Pythia MC simulation and allows one to solve the system of Eqs.~(\ref{eq:forward_central_1}) and~(\ref{eq:forward_central_2}) for $p_q(x)$ and $p_g(x)$ in every \pt and $\eta$ bin separately.
The fractions $f_\text{C}^q$ and $f_\text{F}^q$ obtained from the nominal \Pythia simulation are shown in \cref{fig:fractions} for two \abseta regions.
It is clear that this demixing method relies heavily on the assumed MC simulation and is therefore prone to systematic uncertainties.

\begin{figure}[!htb]
\subfloat[]{
\includegraphics[width=0.49\linewidth]{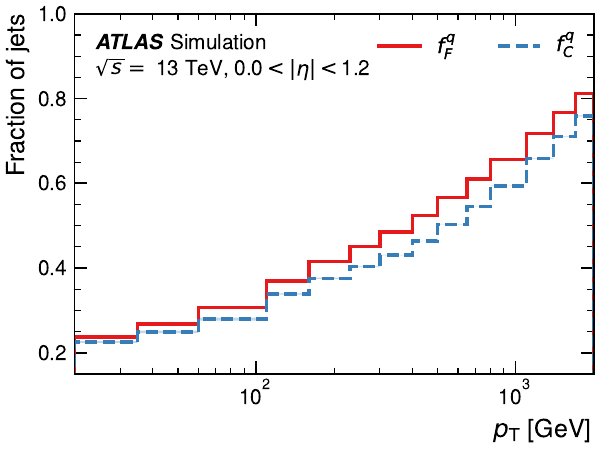}
\label{fig:fractions_ctrl}
}
\subfloat[]{
\includegraphics[width=0.49\linewidth]{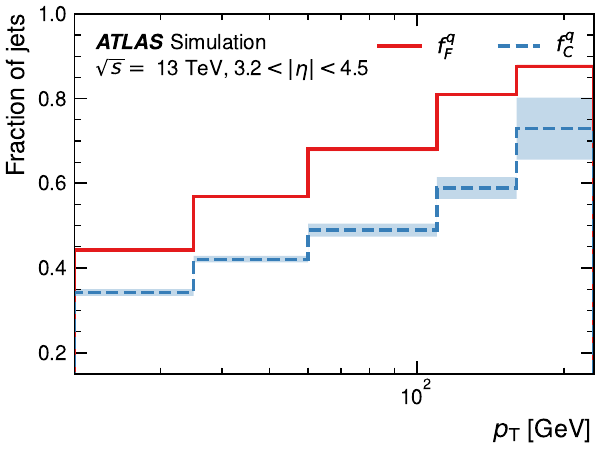}
\label{fig:fractions_fwd}
}
\caption{Fractions of quark jets in the forward ($f_\text{F}^q$) and central ($f_\text{C}^q$) samples as estimated from \Pythia MC simulation for jets with
\protect\subref{fig:fractions_ctrl}~$0.0 < |\eta| < 1.2$ and
\protect\subref{fig:fractions_fwd}~$3.2 < |\eta| < 4.5$.
The error bands represent the statistical uncertainties.
}
\label{fig:fractions}
\end{figure}

\FloatBarrier
\subsection{Jet topics method}
\label{sec:jet_topics}
The second method is the jet topics method~\cite{Metodiev:2018ftz}, which removes the reliance on MC simulations by computing quark and gluon distributions by maximally subtracting the gluon-enriched central distribution from the quark-enriched forward distribution, and vice versa, as
\begin{align}
p_{q}(x) &= \frac{p_\text{F}(x) - \kappa(\text{F}|\text{C}) p_\text{C}(x)}{1 - \kappa(\text{F}|\text{C})}\label{eq:topics_q}\ , \\
p_{g}(x) &= \frac{p_\text{C}(x) - \kappa(\text{C}|\text{F}) p_\text{F}(x)}{1 - \kappa(\text{C}|\text{F})}\label{eq:topics_g}\ .
\end{align}
The so-called reducibility factors $\kappa(\text{F}|\text{C})$ and $\kappa(\text{C}|\text{F})$ are given by
\begin{align}
\kappa(\text{F}|\text{C}) &= \min_x \frac{p_\text{F}(x)}{p_\text{C}(x)} \ , \label{eq:kappa_f_c} \\
\kappa(\text{C}|\text{F}) &= \min_x \frac{p_\text{C}(x)}{p_\text{F}(x)} \ . \label{eq:kappa_c_f}
\end{align}
To minimize the statistical uncertainty in calculating the factors, the $p_\text{F}(x)$ and $p_\text{C}(x)$ distributions are rebinned from an initial uniform binning scheme of 200 bins into a variable-width binning configuration of 20 bins, where the bin boundaries are optimized such that the summed distribution $p_\text{F}(x)+p_\text{C}(x)$ exhibits an approximately uniform population across all bins~\cite{Komiske:2022vxg}.

The version of the jet topics method summarized in Eqs.~(\ref{eq:topics_q}) to~(\ref{eq:kappa_c_f})  requires the quark tagger score distributions to not be a mixture of the gluon ones and some other distributions and vice versa, an assumption referred to as \enquote{mutual irreducibility}.
This condition is not fully satisfied in this analysis and hence a correction is needed.
In this modified version, the maximal subtraction yields a gluon-subtracted quark distribution $p_{q|g}(x)$ and a quark-subtracted gluon distribution $p_{g|q}(x)$ instead:
\begin{align*}
p_{q|g}(x) &= \frac{p_\text{F}(x) - \kappa(\text{F}|\text{C}) p_\text{C}(x)}{1 - \kappa(\text{F}|\text{C})}\ , \\   %
p_{g|q}(x) &= \frac{p_\text{C}(x) - \kappa(\text{C}|\text{F}) p_\text{F}(x)}{1 - \kappa(\text{C}|\text{F})}\ .   %
\end{align*}
The quark and the gluon distributions are then given by
\begin{align*}
p_{q}(x) &= \frac{p_{q|g}(x)(1-\kappa(q|g)) + (1 - \kappa(g|q))\kappa(q|g) p_{g|q}(x)}{1 - \kappa(g|q)\kappa(q|g)}\ , \\   %
p_{g}(x) &= \frac{p_{g|q}(x)(1-\kappa(g|q)) + (1 - \kappa(q|g))\kappa(g|q) p_{q|g}(x)}{1 - \kappa(q|g)\kappa(g|q)}\ ,   %
\end{align*}
where $\kappa(q|g)$ and $\kappa(g|q)$ are the reducibility factors for the quark and gluon distributions \cite{Metodiev:2018ftz}.
The mutual irreducibility condition is satisfied when $\kappa(q|g) = \kappa(g|q) = 0$, indicating no mixing of the distributions.
For each jet \pt and $\eta$ bin, this condition was evaluated by computing the reducibility factors from MC simulations, using formulas similar to Eqs.~(\ref{eq:kappa_f_c}) and~(\ref{eq:kappa_c_f}).
In the majority of kinematic bins, the condition was violated with non-zero reducibility factors, necessitating MC-based corrections obtained by analytically computing $\kappa(q|g)$ and $\kappa(g|q)$.
The impact of such corrections is visible in \cref{fig:extraction}, where the extracted quark and gluon score distributions for the matrix and jet topic methods are compared.

\begin{figure}[!htb]
\subfloat[]{
\includegraphics[width=0.49\linewidth]{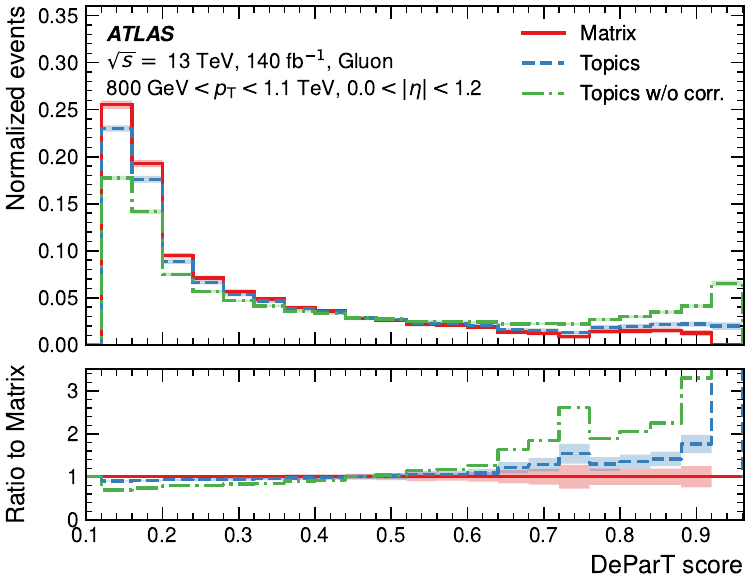}
\label{fig:extraction_g}
}
\subfloat[]{
\includegraphics[width=0.49\linewidth]{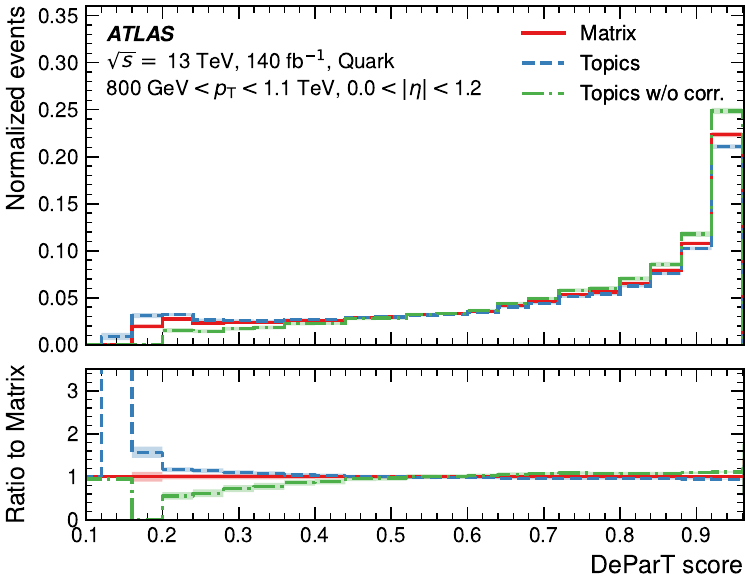}
\label{fig:extraction_q}
}
\caption{The \protect\subref{fig:extraction_g} gluon and \protect\subref{fig:extraction_q} quark DeParT score distributions extracted from Run-2 data with the matrix method (solid red), jet topics without MC correction (dash dotted green), and jet topics with MC correction (dashed blue).
The bottom panels show ratios of jet topics to matrix method results.
The error bands represent statistical uncertainties.
}
\label{fig:extraction}
\end{figure}


%
\FloatBarrier
\section{Tagging efficiency measurement strategy}
\label{sec:sf_def}
%

The extracted quark and gluon score distributions are used to measure the performance of the tagger.
In the following, the score distributions $p_q(x)$ and $p_g(x)$ are normalized to unity.
The quark efficiency is computed from the quark score distribution according to
\begin{equation*}
\varepsilon_q(x_{\mathrm{WP}}) = \int_{x_{\mathrm{WP}}}^1 p_q(x) \, dx,
\end{equation*}
where $p_q(x)$ represents the quark score distribution and $x_{\mathrm{WP}}$ denotes the score threshold defining the working point.
The working point $x_{\mathrm{WP}}$ is determined by requiring a fixed quark efficiency value in the nominal Pythia sample within each jet $\pt$ and $\eta$ bin.
Working points corresponding to fixed quark efficiencies of 50\%, 60\%, 70\%, and 80\% were investigated, with their respective scale factors demonstrating similar characteristics.
The gluon efficiency is calculated from the gluon score distribution as
\begin{equation*}
\varepsilon_g(x_{\mathrm{WP}}) = \int_{x_{\mathrm{WP}}}^1 p_g(x) \, dx,
\end{equation*}
where $p_g(x)$ is the gluon score distribution.

Discrepancies between the tagging efficiencies measured in experimental data and those predicted by MC simulation are quantified through data-to-MC scale factors (SF) computed for each $\pt$ and $\eta$ bin at a specified working point $x_{\mathrm{WP}}$.
The quark efficiency scale factor, which corrects the number of quark jets tagged as quarks, is defined as the ratio of data to MC quark efficiencies:
\begin{equation*}
\mathrm{SF}_q = \frac{\varepsilon_q^{\mathrm{data}}}{\varepsilon_q^{\mathrm{MC}}}.
\end{equation*}
The gluon efficiency scale factor, defined analogously to correct the number of gluon jets tagged as gluons, is expressed as the ratio of data to MC $1 - \varepsilon_g$ values:
\begin{equation*}
\mathrm{SF}_g = \frac{1 - \varepsilon_g^{\mathrm{data}}}{1 - \varepsilon_g^{\mathrm{MC}}}.
\end{equation*}
This definition takes into account that gluon jets constitute a minority of the jet population at the working points considered and hence the gluon scale factor serves primarily as a measure of background efficiency.
The measurements of the 50\%-efficiency working-point scale factors are presented in \cref{sec:results}.


%
\FloatBarrier
\section{Systematic uncertainties}
\label{sec:uncerts}
%

The SF measurements are affected by multiple sources of systematic uncertainty, with contributions which are shared by the matrix and jet topics approaches and others which affect the two methods individually.
Both methodologies are subject to experimental uncertainties arising from the \pileup reweighting, the calibration of the jet energy scale (JES) and jet energy resolution (JER)~\cite{JETM-2018-05}, and the JVT and fJVT SF uncertainties~\cite{PERF-2014-03,PERF-2016-06}.
Theory uncertainties in the MC simulation also affect both models, but their impact is smaller for the jet topic method because that approach depends on MC simulation only through corrections to data-driven estimates.
Additionally, methodology-specific uncertainties are considered.
These include uncertainties associated with the reweighting procedure and the residual MC non-closure, which affect both extraction methods, and an uncertainty originating from the determination of the reducibility factors, which only affects the jet topics method.

\FloatBarrier
\subsection{Experimental uncertainty}
Experimental uncertainties come from three sources: the JES/JER calibration, the JVT/fJVT SFs, and the \pileup reweighting.
The impact of other sources of systematic uncertainty, such as the tracks and PFO identification, was found to be negligible~\cite{JETM-2023-06}.
The JES uncertainty includes components related to the in situ calibration of jets and \pileup, and its extrapolation to higher transverse momentum.
The JER uncertainty is estimated with the dijet balance and random cones measurement methods.
A SF is obtained for each JES/JER variation, and its deviation from the nominal SF value is taken as the systematic uncertainty.
The uncertainty from the reweighting of the MC \pileup distribution to match the distribution in data is evaluated by varying the visible inelastic cross section by its uncertainty.

The statistical uncertainty of the measurements is evaluated using a bootstrapping procedure, wherein each data or MC event is assigned a random weight drawn from a Poisson distribution with unit mean (applied in addition to the nominal weight assigned to a MC event), and the standard deviation of the resulting SFs across 100 independent replicas is taken as the statistical uncertainty estimate.

\FloatBarrier
\subsection{Theory uncertainty}
The theory uncertainty arises from several sources.
The uncertainty from the PS modelling is estimated by comparing SFs obtained with the same ME and hadronization model but different shower algorithms.
In the case of Run-2 data, the envelope of the SFs obtained with the Pythia, Herwig String, and Sherpa String samples is taken as the PS uncertainty.
Using the Herwig String sample, which employs an untuned Lund string fragmentation model without colour-reconnection effects, yields a conservative uncertainty estimate.
The Run-3 PS uncertainty is estimated from the difference between the SFs obtained with the Herwig Angular and Herwig Dipole samples.
This was made possible by the availability of new MC samples developed for Run-3 conditions.
In recent measurements of the Lund subjet multiplicities~\cite{STDM-2023-07}, which are observables sensitive to the PS modelling and relevant to $q/g$ tagging, discrepancies of more than 50\% between various MC generators and data were seen.
This is consistent with the present analysis finding the PS uncertainty to be the dominant theory uncertainty.

The impact of initial-state radiation uncertainties is evaluated by varying the Var3c set of parameters in the A14 tune of the PS~\cite{ATL-PHYS-PUB-2014-021,ATL-PHYS-PUB-2017-007}.
Variations of the renormalization (\muR) and factorization (\muF) scales for initial- and final-state radiation are used to estimate the uncertainty due to missing higher-order corrections in the PS modelling.
Seven possibilities for (\muR,\,\muF), where their nominal values are multiplied by factors of (0.5,\,0.5), (0.5,\,1), (1,\,0.5), (1,\,1), (2,\,1), (1,\,2) and (2,\,2), are used to estimate an envelope of SF values for the Pythia sample.
The width of this envelope is taken as the associated uncertainty.

Splitting-kernel variations~\cite{Mrenna:2016sih} are also propagated to alternative SF measurements, and the uncertainty associated with the non-singular part of the splitting functions for the initial- and final-state radiation is taken as the envelope of those SF values.

The systematic uncertainty due to the hadronization modelling is estimated as the difference between the SFs obtained with the Sherpa String and Sherpa Cluster samples for both Run-2 and Run-3 data.
An additional uncertainty covering differences in the ME calculation, and its matching to the PS algorithm, is estimated from the differences between the SFs obtained by using the Powheg+Pythia and nominal Pythia samples.
The uncertainty due to the PDF is evaluated in accord with the LHAPDF recommendations~\cite{Buckley:2014ana} by using variations of the \NNPDF[2.3lo] PDF set in the nominal Pythia sample.

\FloatBarrier
\subsection{Methodology uncertainty}
Both the matrix and jet topics methods suffer from uncertainties associated with the central-to-forward reweighting process and with the residual MC non-closure.
To estimate the systematic uncertainty from the reweighting process, the nominal reweighting choice described in \cref{sec:extraction} is varied by using either the quark or gluon jet distributions to derive the weights, and by applying the correction in either direction, matching $p_\text{C}(x)$ to $p_\text{F}(x)$ or matching $p_\text{F}(x)$ to $p_\text{C}(x)$.
The SF envelope obtained from these four alternative reweighting configurations is taken as the systematic uncertainty.

The MC non-closure uncertainty is obtained from the difference between the truth-level particle distributions obtained directly from simulation and the distributions extracted via the matrix method or jet topics procedure when applied to the same MC samples.
Additionally, the nominal SFs and efficiencies are shifted by the MC non-closure.

The jet topics method uncertainties arise from the calculation of the reducibility factors.
As explained in \cref{sec:extraction}, the calculation is based on finding the minimum of a distribution and suffers from significant statistical fluctuations.
Two calculation methods, described in Ref.~\cite{Komiske:2022vxg}, are used to estimate the uncertainty: the quantiled log-likelihood method with a rebinning into 10 bins and a degree~2 polynomial fit to find the minimum, and the ROC fit method with a degree~2 polynomial.
The envelope of the SFs obtained with the two methods is taken as the uncertainty associated with the reducibility factors computation.


%
\FloatBarrier
\section{Results}
\label{sec:results}
%

The measured DeParT tagger efficiencies, extracted via the jet topics method from Run-2 data, are compared with MC predictions from the nominal Pythia sample in \cref{fig:eff}.
The quark and gluon efficiencies are evaluated at a fixed working point corresponding to 50\% quark efficiency in the MC simulation and their values and statistical uncertainties are smoothed.
The smoothing employs Gaussian kernels as in Ref.~\cite{JETM-2018-05}, applied to 100 bootstrapped resamples of the efficiency distribution.
For each resample, bin contents are drawn from Gaussian distributions with means and standard deviations equal to the nominal bin contents and their uncertainties.
The smoothed value and its uncertainty are determined from the mean and standard deviation of the 100 smoothed distributions, respectively.

In the central pseudorapidity region ($0.0 < |\eta| <  1.2$), the $\varepsilon_q$ measurements (\cref{fig:eff_quark_run2_ctrl}) reveal a characteristic systematic shift of approximately 2--3\% relative to MC predictions throughout the intermediate transverse momentum range of 200~\gev to 1~\tev.
At \pt values below 200~\gev the measured $\varepsilon_q$ agree with MC expectations within statistical uncertainties.
In the forward pseudorapidity region with $3.2 < |\eta| < 4.5$, the measured $\varepsilon_q$ values (\cref{fig:eff_quark_run2_fwd}) are approximately 5--6\% lower than the MC predictions across the entire \pt spectrum.
The $1-\varepsilon_g$ data measurements in the central region (\cref{fig:eff_gluon_run2_ctrl}) are consistently lower than the MC predictions between 20~\gev and 200~\gev.
Between 200~\gev and 800~\gev the measured $1-\varepsilon_g$ agree with MC expectations within statistical uncertainties.
Above 800~\gev, the measured $1-\varepsilon_g$ are systematically lower than MC expectations, signalling potential deficiencies in the modelling of high-\pt gluon jet properties.
The forward region $1-\varepsilon_g$ measurements (\cref{fig:eff_gluon_run2_fwd}) exhibit a systematic downward shift with respect to the MC predictions across the \pt spectrum.

\begin{figure}[!htb]
\subfloat[]{
\includegraphics[width=0.49\linewidth]{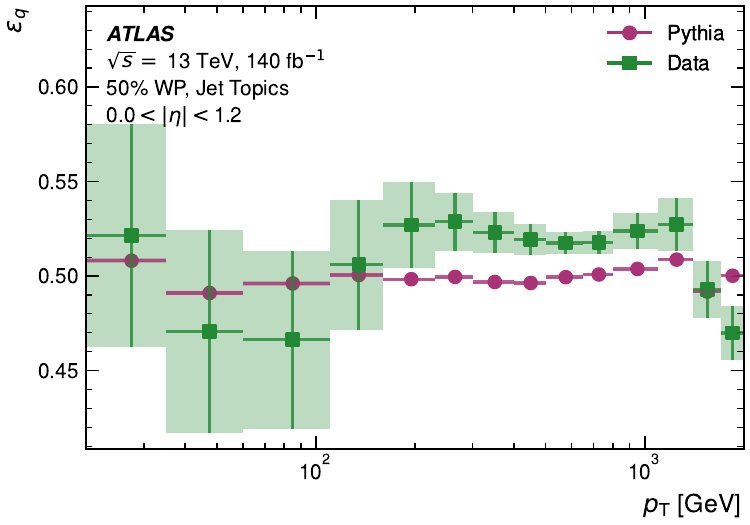}
\label{fig:eff_quark_run2_ctrl}
}\subfloat[]{
\includegraphics[width=0.49\linewidth]{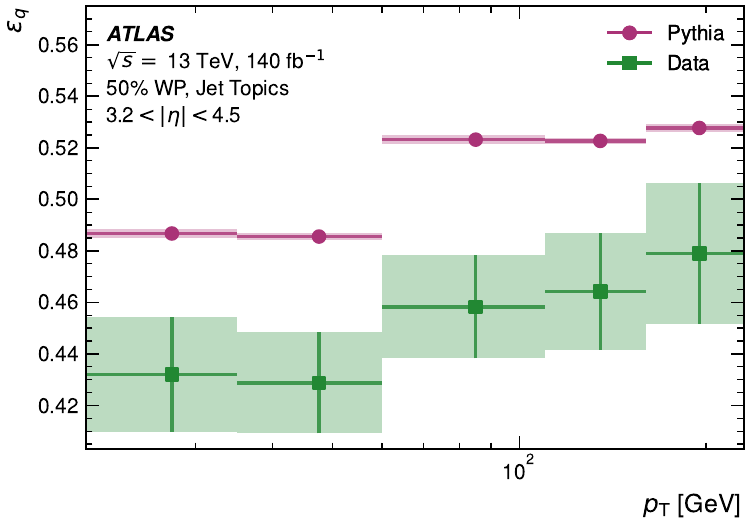}
\label{fig:eff_quark_run2_fwd}
}\\
\subfloat[]{
\includegraphics[width=0.49\linewidth]{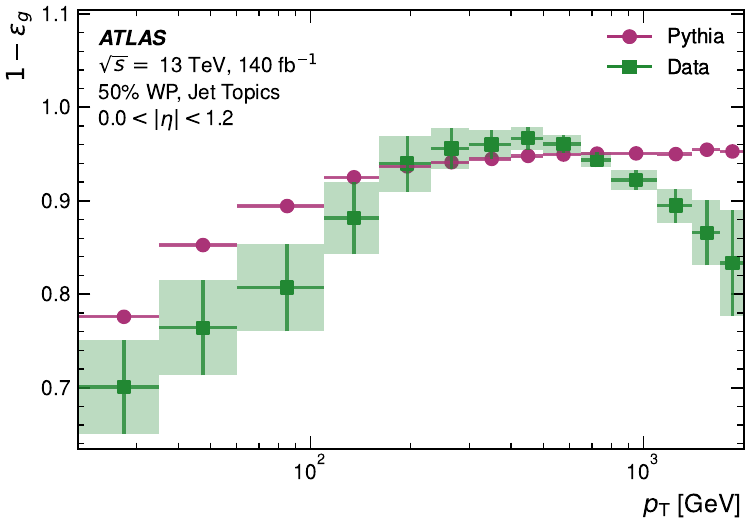}
\label{fig:eff_gluon_run2_ctrl}
}\subfloat[]{
\includegraphics[width=0.49\linewidth]{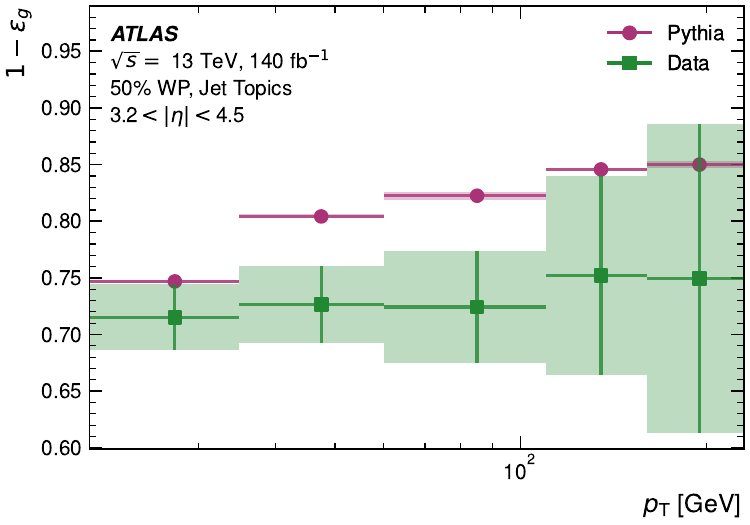}
\label{fig:eff_gluon_run2_fwd}
}
\caption{The MC efficiencies for truth-labelled
\protect\subref{fig:eff_quark_run2_ctrl}~quark jets with $0.0 < |\eta| <  1.2$,
\protect\subref{fig:eff_quark_run2_fwd}~quark jets with $3.2 < |\eta| < 4.5$,
\protect\subref{fig:eff_gluon_run2_ctrl}~gluon jets with $0.0 < |\eta| <  1.2$, and
\protect\subref{fig:eff_gluon_run2_fwd}~gluon jets with $3.2 < |\eta| < 4.5$,
compared with the efficiencies measured for Run-2 data with the jet topics method.
Gluon efficiencies are reported as $1-\varepsilon_g$ to ease their comparison with the gluon SF measurements.
The 50\% WP corresponds to a score threshold defined by a fixed 0.5 quark efficiency in the Pythia sample in each \pt and \abseta bin.
The deviations of the truth quark-jet efficiencies from 0.5 are due to the finite binning of the DeParT score used in the efficiency calculation.
Only the statistical uncertainties are shown.
}
\label{fig:eff}
\end{figure}

\begin{figure}[!htb]
\centering
\includegraphics[width=1.\linewidth]{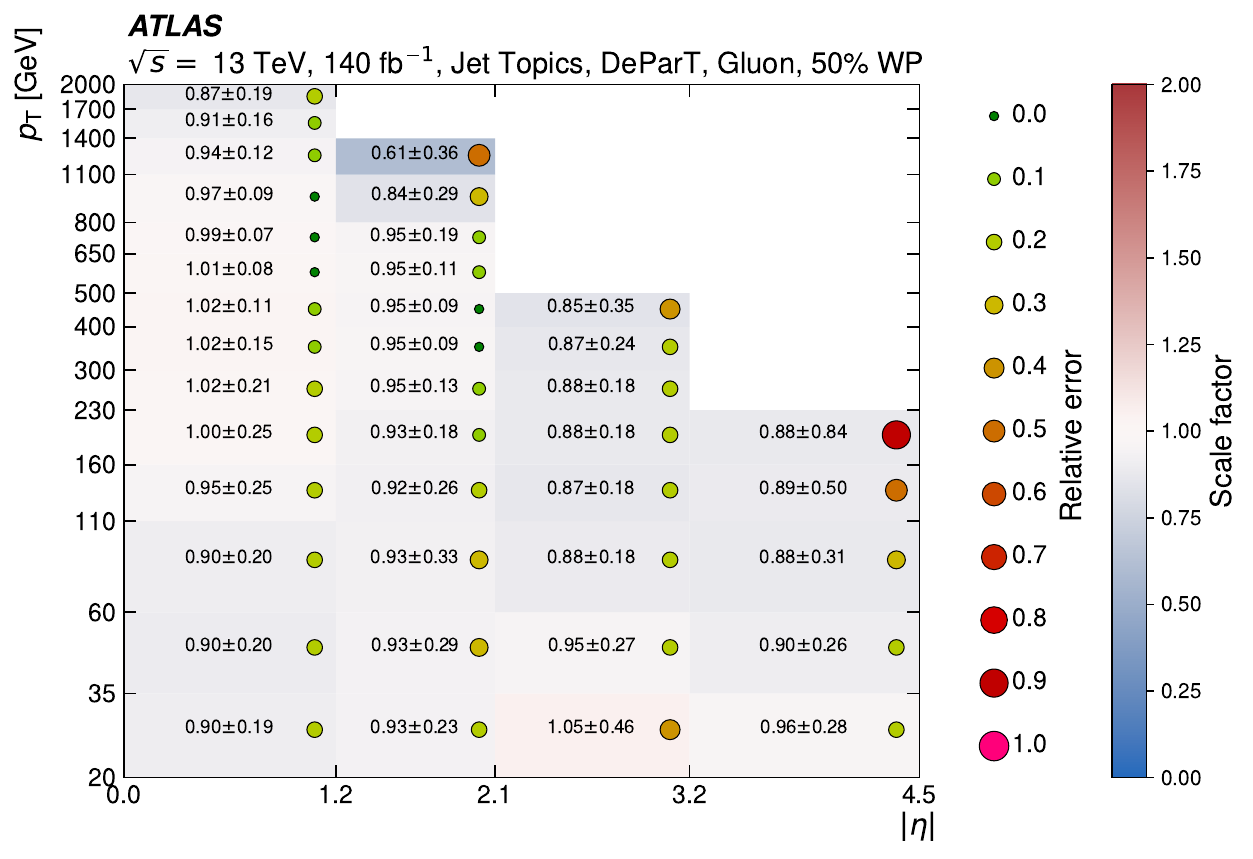}
\caption{The gluon efficiency scale factors derived with the jet topics method in each \pt and \abseta bin for Run-2 jets.
The size and the colour of the dots represents the relative total uncertainty of the scale factor.
The 50\% WP corresponds to a score threshold defined by a fixed 0.5 quark efficiency in the Pythia sample in each \pt and \abseta bin.
The nominal scale factor's value and total uncertainty is shown in each bin.
}
\label{fig:sf_gluon_eff_run2}
\end{figure}
\begin{figure}[!htb]
\centering
\includegraphics[width=1.\linewidth]{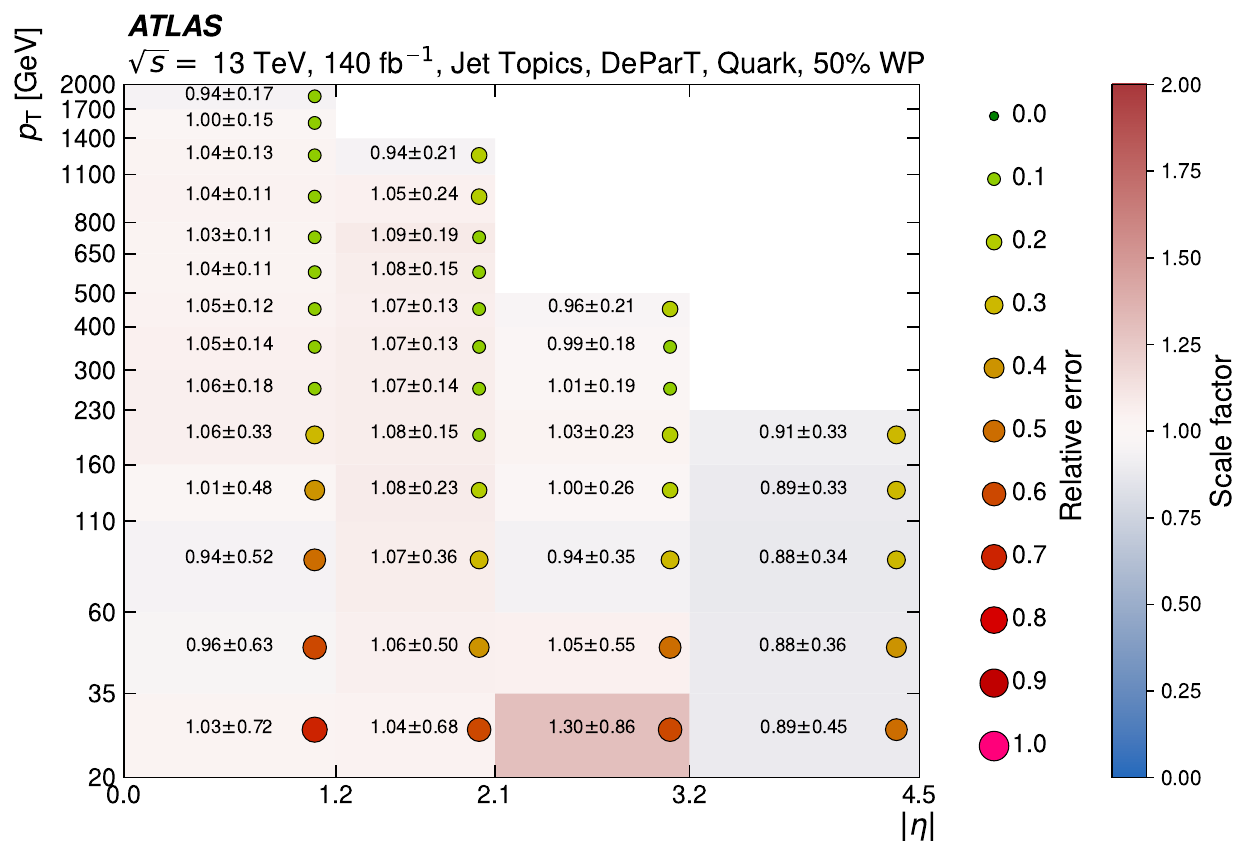}
\caption{The quark efficiency scale factors derived with the jet topics method in each \pt and \abseta bin for Run-2 jets.
The size and the colour of the dots represents the relative total uncertainty of the scale factor.
The 50\% WP corresponds to a score threshold defined by a fixed 0.5 quark efficiency in the Pythia sample in each \pt and \abseta bin.
The nominal scale factor's value and total uncertainty is shown in each bin.
}
\label{fig:sf_quark_eff_run2}
\end{figure}

The SFs measured using the jet topics method and the Run-2 data are presented for gluon and quark jets in their full kinematic phase space in \cref{fig:sf_gluon_eff_run2,fig:sf_quark_eff_run2}, respectively.
Individual components of the total uncertainty are smoothed using a Gaussian kernel without bootstrapping as is done in Ref.~\cite{JETM-2018-05}, while the nominal SF values are smoothed using a bootstrapped Gaussian kernel as described above for the efficiencies.
The gluon SF values (\cref{fig:sf_gluon_eff_run2}) are generally consistent with unity within the total uncertainty across the measured phase space.
The nominal SF values range from 0.61 to 1.05.
Similarly, the quark SFs (\cref{fig:sf_quark_eff_run2}) are also consistent with unity, with values ranging from 0.88 to 1.30 and the total relative uncertainty becoming larger for low \pt and reaching about 70\%.

\begin{figure}[!htb]
\subfloat[]{
\includegraphics[width=0.49\linewidth]{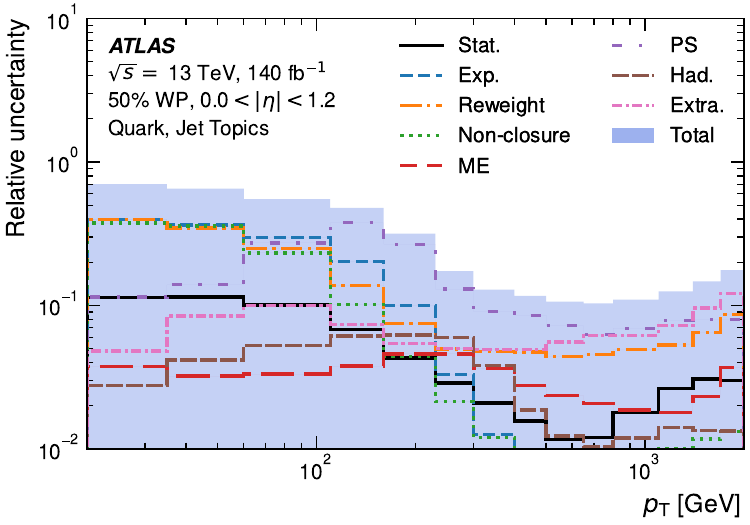}
\label{fig:uncerts_quark_run2_ctrl}
}\subfloat[]{
\includegraphics[width=0.49\linewidth]{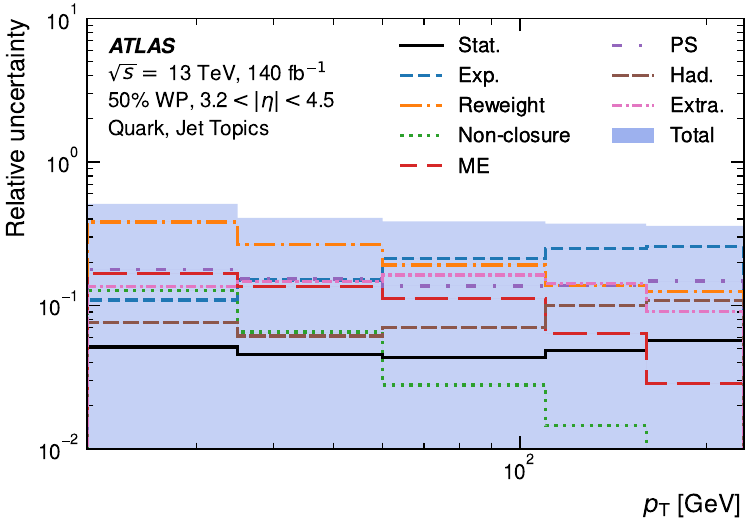}
\label{fig:uncerts_quark_run2_fwd}
}\\
\subfloat[]{
\includegraphics[width=0.49\linewidth]{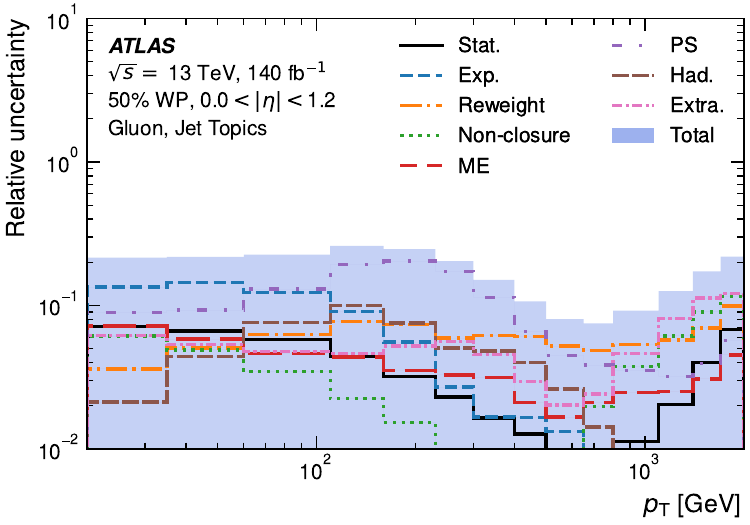}
\label{fig:uncerts_gluon_run2_ctrl}
}\subfloat[]{
\includegraphics[width=0.49\linewidth]{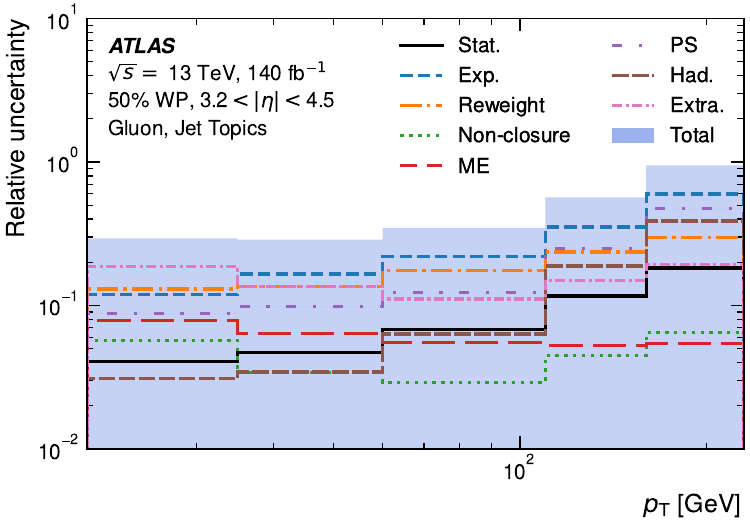}
\label{fig:uncerts_gluon_run2_fwd}
}
\caption{The scale factor uncertainties derived with the jet topics method for Run-2
\protect\subref{fig:uncerts_quark_run2_ctrl}~quark jets with $0.0 < |\eta| <  1.2$,
\protect\subref{fig:uncerts_quark_run2_fwd}~quark jets with $3.2 < |\eta| < 4.5$,
\protect\subref{fig:uncerts_gluon_run2_ctrl}~gluon jets with $0.0 < |\eta| <  1.2$, and
\protect\subref{fig:uncerts_gluon_run2_fwd}~gluon jets with $3.2 < |\eta| < 4.5$.
The 50\% WP corresponds to a score threshold defined by a fixed 0.5 quark efficiency in the Pythia sample in each \pt and \abseta bin.
The PDF uncertainty is not shown, as it is negligible in the considered phase space.
The PS component stands for the quadrature sum of the following components: comparison of PS modelling by different MC generators, Var3c parameter variations, \muR and \muF scale variations, and splitting-kernel variations.
The theory uncertainty due to hadronization modelling is denoted as `Had.' and the systematic uncertainty arising from the calculation of the reducibility factors, the extraction uncertainty, is denoted as `Extra.'.
The total uncertainty is represented by the blue shaded band, with individual systematic components shown as coloured lines.
}
\label{fig:uncerts}
\end{figure}

A breakdown of the uncertainties affecting the SFs measured using the jet topics method for Run-2 data is shown in \cref{fig:uncerts} for both the quark and gluon jets in the central and forward \abseta regions.
Both the quark and gluon SFs in the $0.0 < |\eta| <  1.2$ region are dominated by the PS uncertainty between approximately 100~\gev and 500~\gev and by methodological uncertainties at \pt above 1~\tev.
The low-\pt region is dominated by both experimental and methodological uncertainties for quark SFs and by experimental uncertainties for gluon SFs.
In the $3.2 < |\eta| < 4.5$ region, the reweighting and experimental uncertainties dominate in the quark SF case, while the extraction and experimental uncertainties are the leading contributions for the gluon SFs.
The total relative uncertainty of gluon SFs ranges approximately from 10\% to 25\% in the $0.0 < |\eta| <  1.2$ region and 30\% to 95\% in the $3.2 < |\eta| < 4.5$ region.
The quark SF uncertainties range approximately from 10\% to 70\% in the $0.0 < |\eta| <  1.2$ region and 35\% to 50\% in the $3.2 < |\eta| < 4.5$ region.
The intermediate pseudorapidity region, $1.2 < |\eta| < 3.2$, exhibits similar trends in the total uncertainties, as demonstrated in \cref{fig:sf_gluon_eff_run2,fig:sf_quark_eff_run2}.
Statistical uncertainties become significant at the highest \pt values, where the available sample is much smaller, and also at the lowest \pt values, due to the use of prescaled triggers.
The MC non-closure uncertainty exhibits pronounced \pt dependence, becoming increasingly significant at lower \pt values.
There the PDF properties render the quark and gluon fractions in the forward sample more similar to those in the central sample, thereby reducing the effective discrimination power of the pseudorapidity-based selection strategy and making the extraction procedure numerically unstable.

\begin{figure}[!htb]
\subfloat[]{
\includegraphics[width=0.49\linewidth]{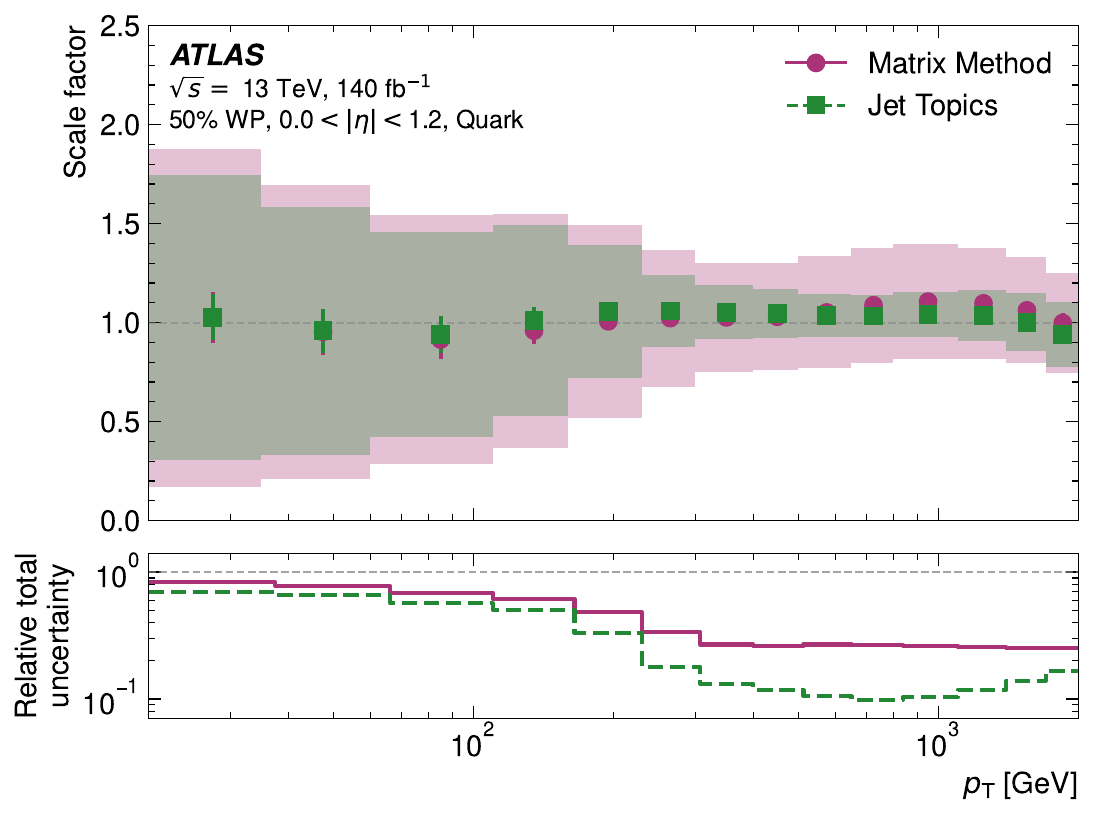}
\label{fig:compare_quark_run2_ctrl}
}\subfloat[]{
\includegraphics[width=0.49\linewidth]{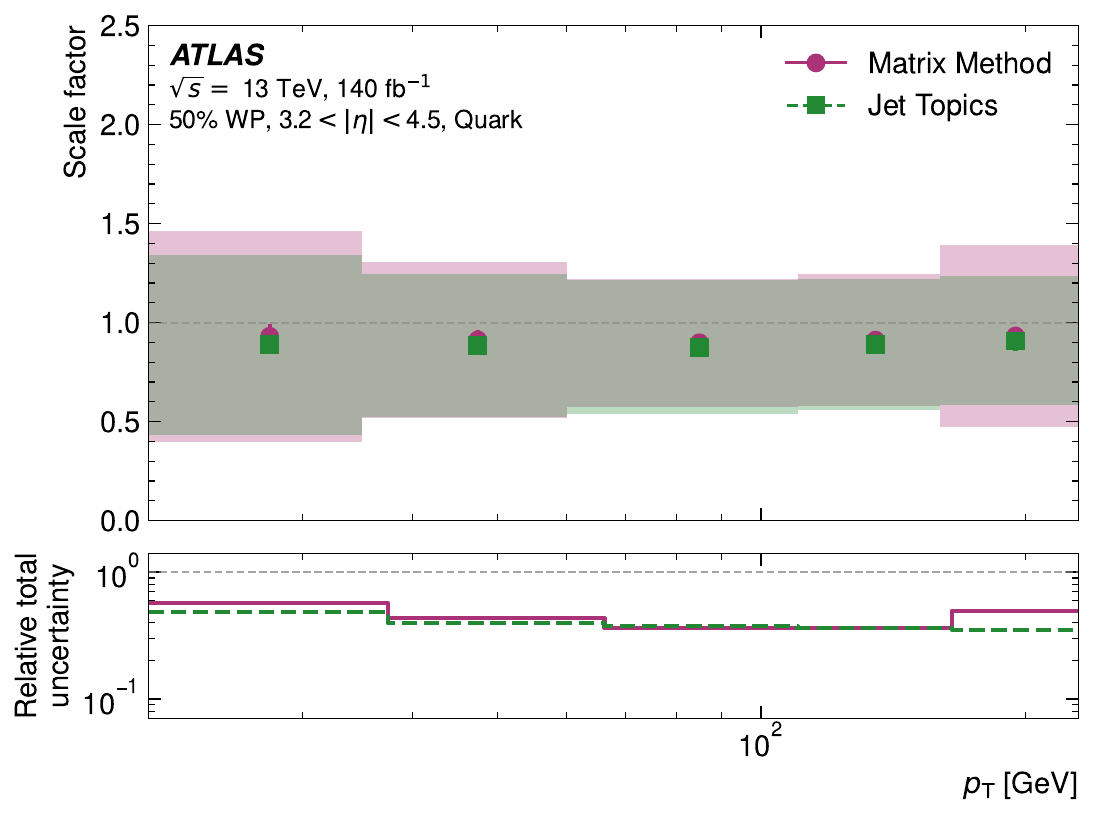}
\label{fig:compare_quark_run2_fwd}
}\\
\subfloat[]{
\includegraphics[width=0.49\linewidth]{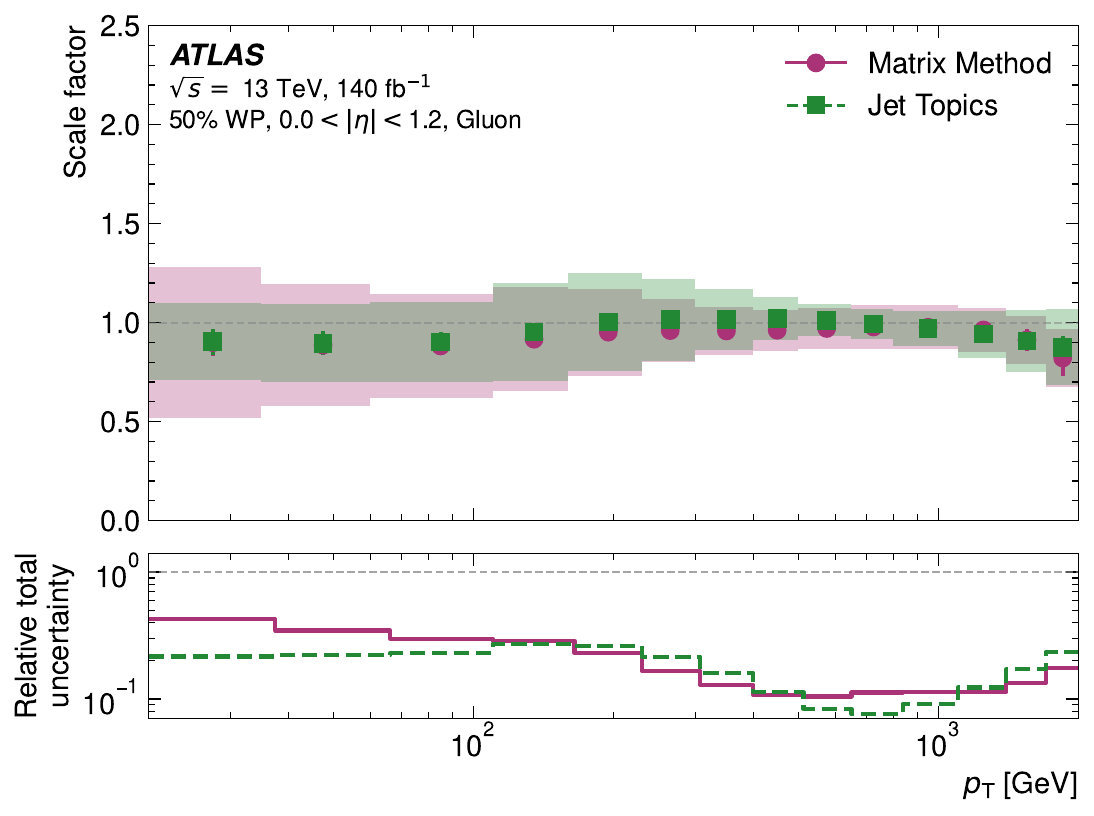}
\label{fig:compare_gluon_run2_ctrl}
}\subfloat[]{
\includegraphics[width=0.49\linewidth]{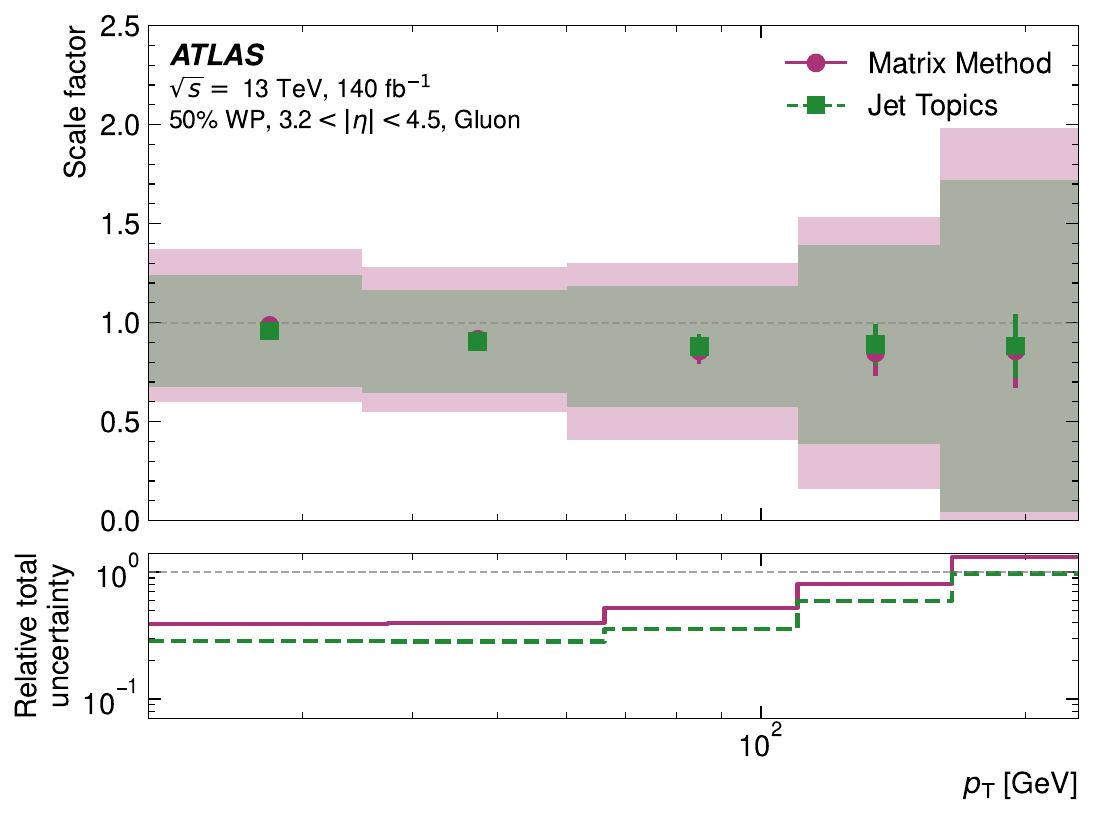}
\label{fig:compare_gluon_run2_fwd}
}
\caption{Comparison of SFs and total uncertainties derived with the jet topics method and matrix method for Run-2
\protect\subref{fig:compare_quark_run2_ctrl}~quark jets with $0.0 < |\eta| <  1.2$,
\protect\subref{fig:compare_quark_run2_fwd}~quark jets with $3.2 < |\eta| < 4.5$,
\protect\subref{fig:compare_gluon_run2_ctrl}~gluon jets with $0.0 < |\eta| <  1.2$, and
\protect\subref{fig:compare_gluon_run2_fwd}~gluon jets with $3.2 < |\eta| < 4.5$.
The 50\% WP corresponds to a score threshold defined by a fixed 0.5 quark efficiency in the Pythia sample in each \pt and \abseta bin.
The error bars represent the statistical uncertainties and the shaded bands represent the total uncertainties.
The bottom panels show the relative total uncertainties for the jet topics and matrix methods.
}
\label{fig:compare}
\end{figure}

A comparison of the scale factors derived via the jet topics and matrix methods is presented in \cref{fig:compare}, revealing distinct precision advantages that vary across kinematic regimes and jet flavours.
The two extraction methodologies demonstrate consistency in their central SF values while exhibiting differences in their associated uncertainties.
For quark jets in the central pseudorapidity region (\cref{fig:compare_quark_run2_ctrl}), the jet topics method reduces the total uncertainties, delivering improvements of up to 15\% in relative uncertainty in the intermediate \pt range of 300--1000~\gev, where theory uncertainties typically dominate.
Although its better precision persists into the forward region (\cref{fig:compare_quark_run2_fwd}), the jet topics method's advantage decreases markedly, likely due to both methods becoming statistically limited by smaller event yields.
The gluon SF measurements reveal a more nuanced uncertainty landscape: in the central region (\cref{fig:compare_gluon_run2_ctrl}) the jet topics method demonstrates superior precision at low \pt, achieving up to 20\% smaller uncertainties in the 20--110~\gev range.
Above 110~\gev, the two methods yield comparable uncertainties.
An improvement also emerges in the forward pseudorapidity region for gluon jets (\cref{fig:compare_gluon_run2_fwd}), where the jet topics method consistently reduces the uncertainties, with improvements of up to 15\%.
The jet topics method achieves these improvements over the matrix method primarily by reducing the PS and ME uncertainties via less dependence on MC simulation, relying on it only for the MC correction described in \cref{sec:jet_topics}.

\begin{figure}[!htb]
\subfloat[]{
\includegraphics[width=0.49\linewidth]{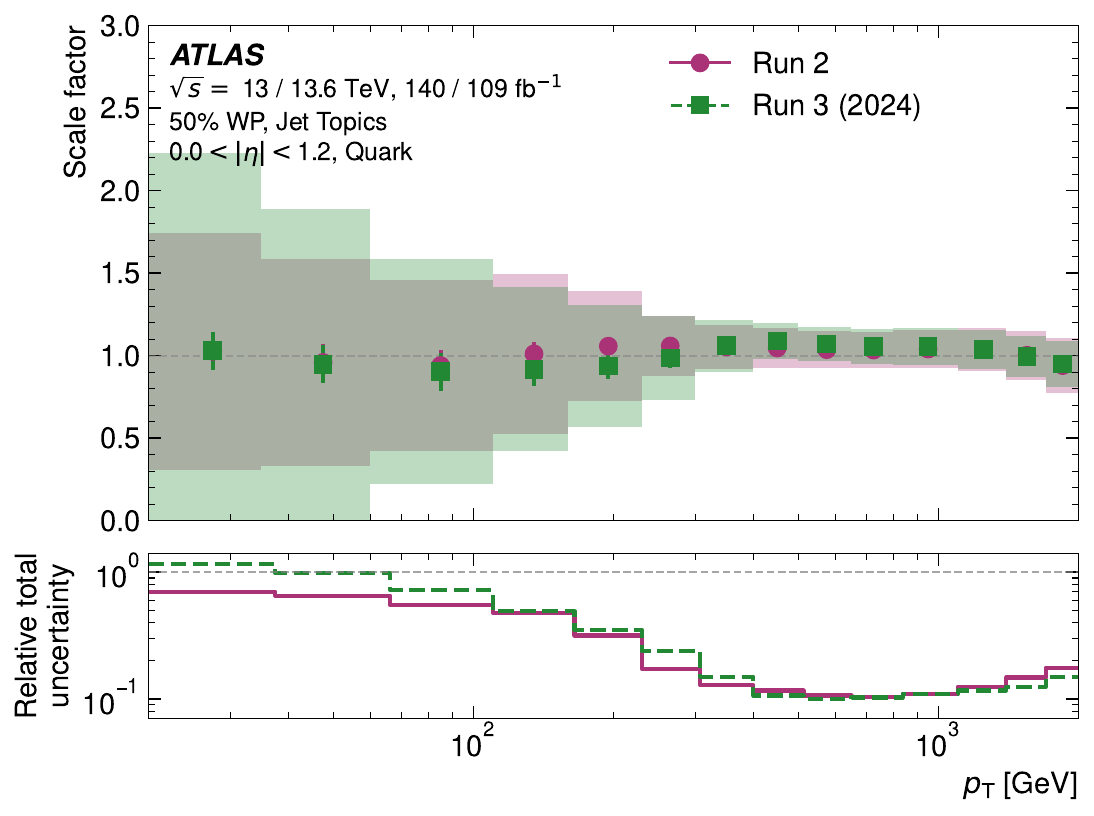}
\label{fig:runs_compare_quark_ctrl}
}\subfloat[]{
\includegraphics[width=0.49\linewidth]{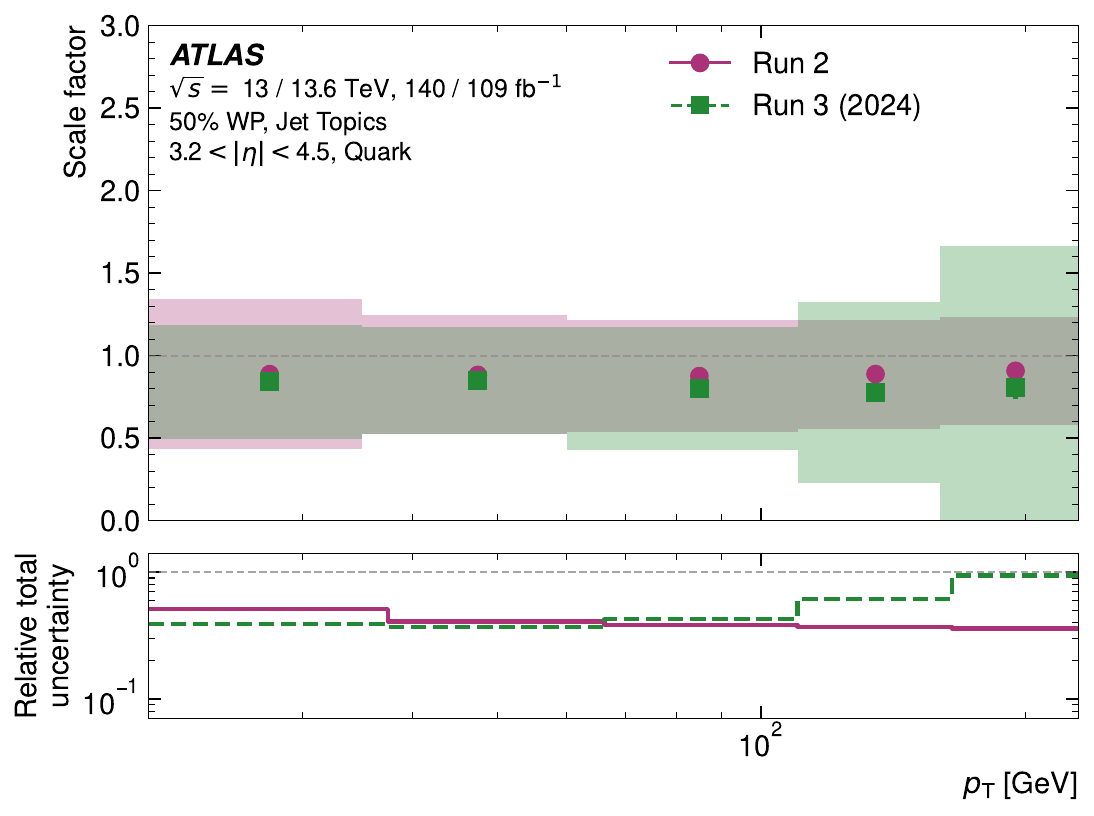}
\label{fig:runs_compare_quark_fwd}
}\\
\subfloat[]{
\includegraphics[width=0.49\linewidth]{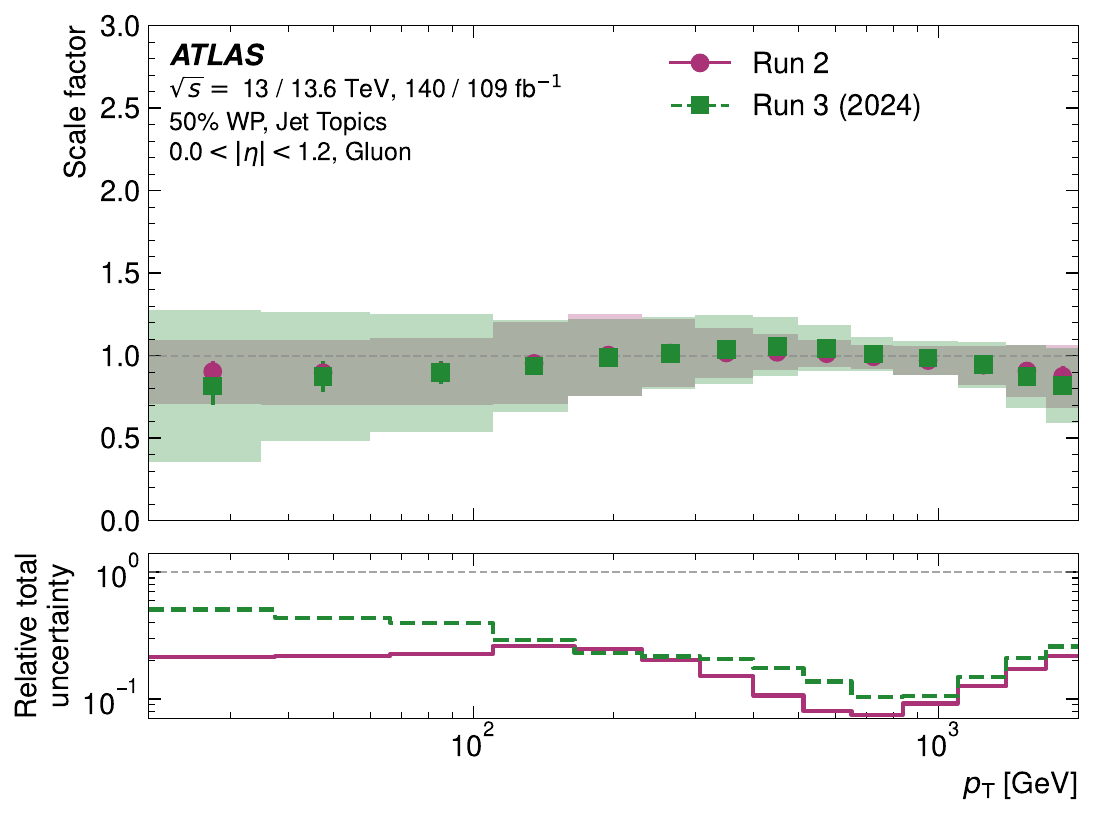}
\label{fig:runs_compare_gluon_ctrl}
}\subfloat[]{
\includegraphics[width=0.49\linewidth]{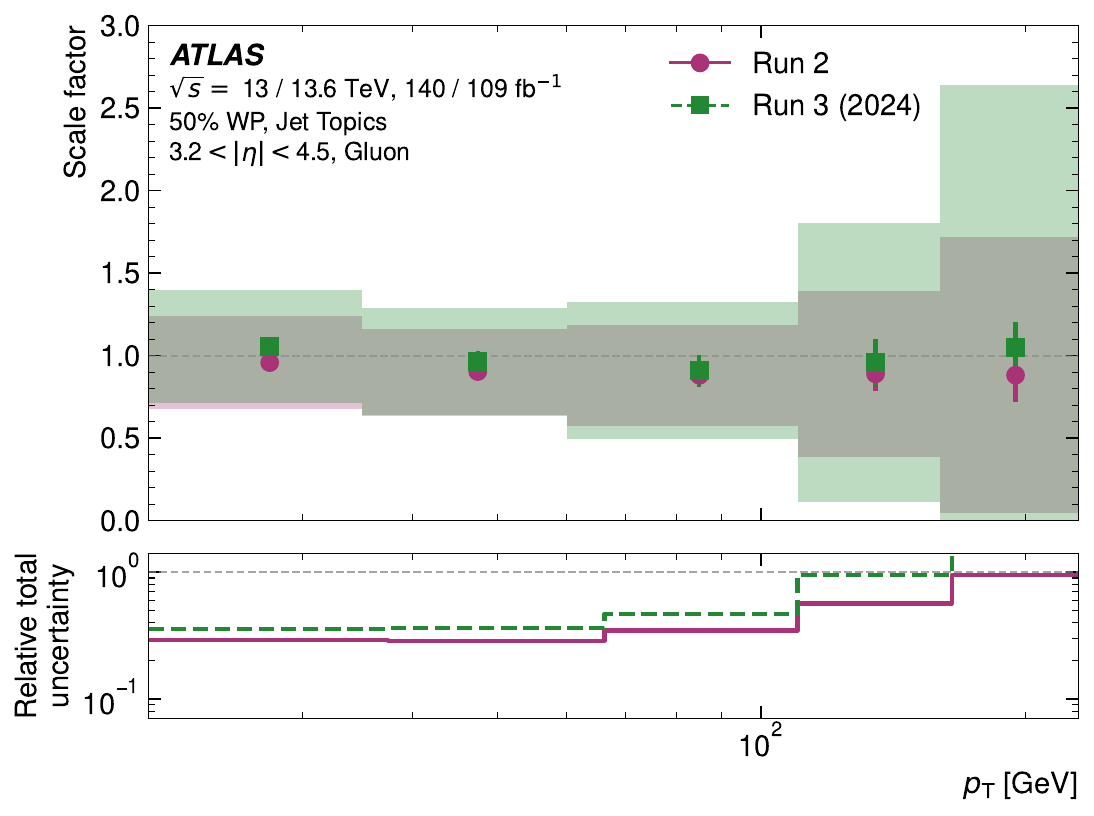}
\label{fig:runs_compare_gluon_fwd}
}
\caption{Comparison of SFs and total uncertainties derived with the jet topics method with Run-2 and Run-3 data for
\protect\subref{fig:runs_compare_quark_ctrl}~quark jets with $0.0 < |\eta| <  1.2$,
\protect\subref{fig:runs_compare_quark_fwd}~quark jets with $3.2 < |\eta| < 4.5$,
\protect\subref{fig:runs_compare_gluon_ctrl}~gluon jets with $0.0 < |\eta| <  1.2$, and
\protect\subref{fig:runs_compare_gluon_fwd}~gluon jets with $3.2 < |\eta| < 4.5$.
The 50\% WP corresponds to a score threshold defined by a fixed 0.5 quark efficiency in the Pythia sample in each \pt and \abseta bin.
The error bars represent the statistical uncertainties and the shaded bands represent the total uncertainties.
The bottom panels show the relative total uncertainties for \RunTwo and \RunThr.
}
\label{fig:runs_compare}
\end{figure}

The SFs derived with the jet topics method for Run-3 data are compared with those from \RunTwo in \cref{fig:runs_compare}.
The differences between the two runs' results arise from several factors: different integrated luminosities, different numbers of simulated events, different \pileup conditions, differences in trigger strategies, and the use of different MC generators for deriving the PS uncertainty.
At low jet \pt, the number of simulated events available for Run-3 conditions is around 30\% of that for Run-2 conditions, much lower than the ratio of the integrated luminosities of the two data-taking periods (around 70\%).
This statistical limitation directly impacts the systematic uncertainties, as the reweighting described in \cref{sec:reweighting} uses weights derived from ratios of MC distributions, whereby statistical fluctuations lead to larger uncertainties in the corrected distributions.
The PS uncertainty in \RunTwo is taken from the envelope of the SFs obtained with the Pythia, Herwig String, and Sherpa String samples, while in \RunThr the uncertainty is estimated from the difference between the SFs obtained with the Herwig Angular and Herwig Dipole samples.
In \RunThr the total uncertainties reach over 100\% for quark SFs at low \pt in the $0.0 < |\eta| <  1.2$ region and for gluon SFs at high \pt in the $3.2 < |\eta| < 4.5$ region.
A detailed breakdown of the relative systematic uncertainties in the SFs measured using the jet topics method in Run-3 data is illustrated in \cref{fig:uncerts_run3} for both the quark and gluon jets in the central and forward \abseta regions.

\begin{figure}[!htb]
\subfloat[]{
\includegraphics[width=0.49\linewidth]{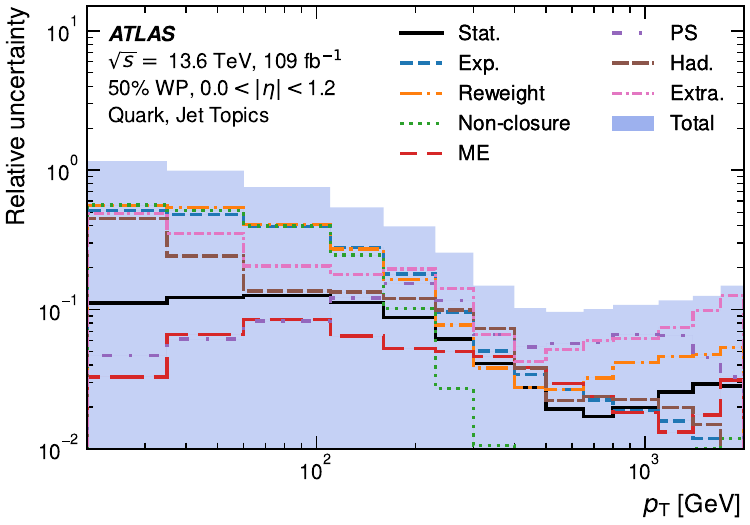}
\label{fig:uncerts_quark_run3_ctrl}
}\subfloat[]{
\includegraphics[width=0.49\linewidth]{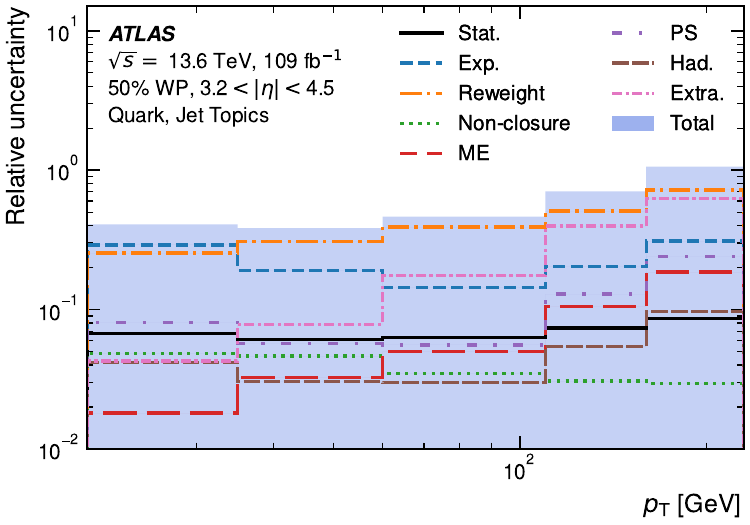}
\label{fig:uncerts_quark_run3_fwd}
}\\
\subfloat[]{
\includegraphics[width=0.49\linewidth]{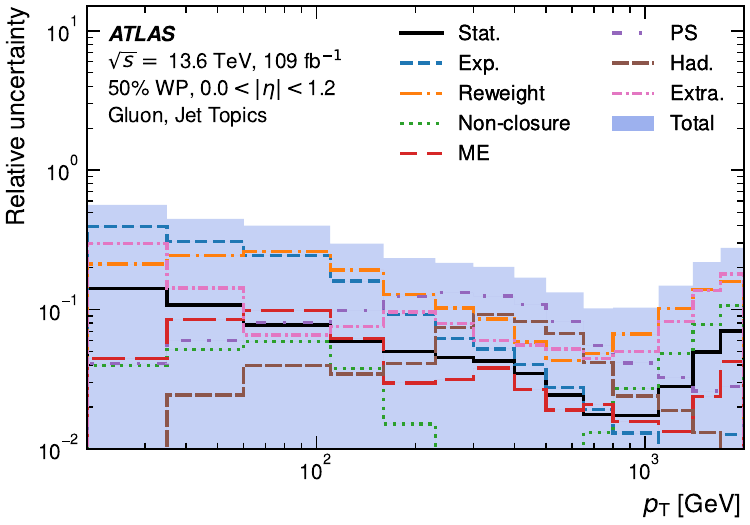}
\label{fig:uncerts_gluon_run3_ctrl}
}\subfloat[]{
\includegraphics[width=0.49\linewidth]{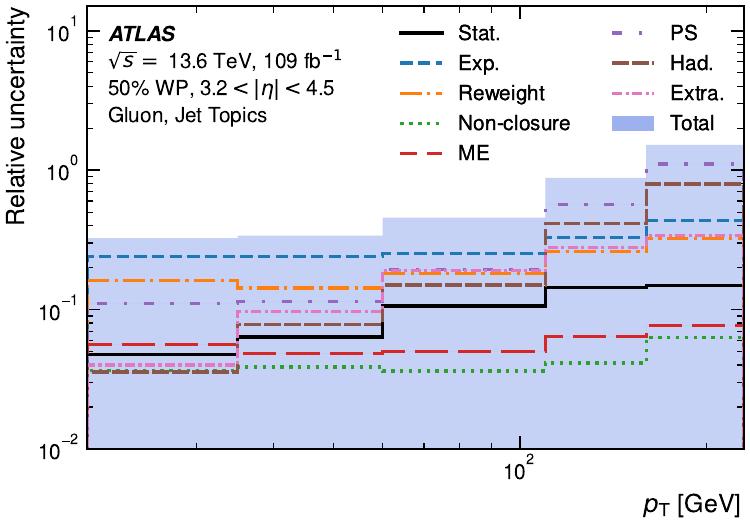}
\label{fig:uncerts_gluon_run3_fwd}
}
\caption{The scale factor uncertainties derived with the jet topics method for Run-3
\protect\subref{fig:uncerts_quark_run3_ctrl}~quark jets with $0.0 < |\eta| <  1.2$,
\protect\subref{fig:uncerts_quark_run3_fwd}~quark jets with $3.2 < |\eta| < 4.5$,
\protect\subref{fig:uncerts_gluon_run3_ctrl}~gluon jets with $0.0 < |\eta| <  1.2$, and
\protect\subref{fig:uncerts_gluon_run3_fwd}~gluon jets with $3.2 < |\eta| < 4.5$.
The 50\% WP corresponds to a score threshold defined by a fixed 0.5 quark efficiency in the Pythia sample in each \pt and \abseta bin.
The PDF uncertainty is not shown, as it is negligible in the considered phase space.
The PS component stands for the quadrature sum of the following components: comparison of PS modelling by different MC generators, Var3c parameter variations, \muR and \muF scale variations, and splitting-kernel variations.
The theory uncertainty due to hadronization modelling is denoted as `Had.' and the systematic uncertainty arising from the calculation of the reducibility factors, the extraction uncertainty, is denoted as `Extra.'.
The total uncertainty is represented by the blue shaded band, with individual systematic components shown as coloured lines.
}
\label{fig:uncerts_run3}
\end{figure}

Since analyses interested in using the results reported above may use different MC samples, a MC-to-MC SF is obtained by using each of the alternative MC samples in the SF definition and treating the Pythia samples as pseudo-data, to account for modelling differences between the Pythia and alternative MC samples.
The MC-to-MC SFs are shown in \cref{fig:mc-to-mc} for quark and gluon jets in the central and forward regions.
In the central region, the quark MC-to-MC SF (\cref{fig:mc-to-mc_quark_ctrl}) has a relatively large value above unity for the Herwig String sample, which is inconsistent with all the other samples.
For the gluon MC-to-MC SF (\cref{fig:mc-to-mc_gluon_ctrl}), all the samples diverge from unity for small \pt values, in both the forward and central regions, with the Sherpa Cluster sample being the furthest from unity.

\begin{figure}[!htb]
\subfloat[]{
\includegraphics[width=0.49\linewidth]{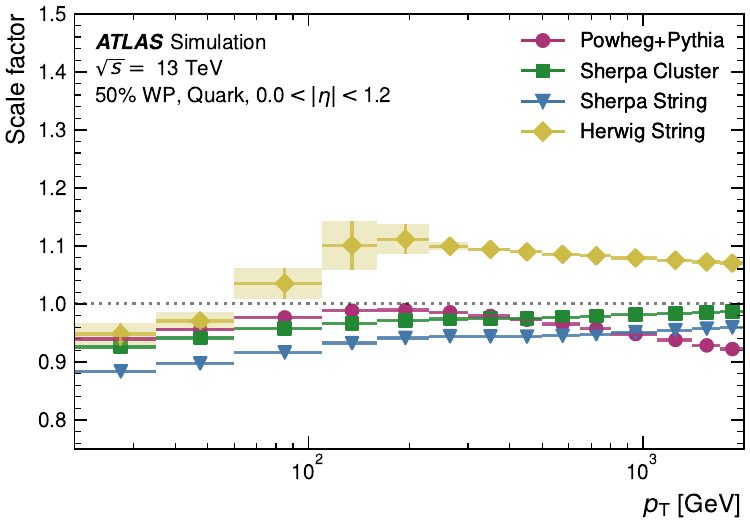}
\label{fig:mc-to-mc_quark_ctrl}
}\subfloat[]{
\includegraphics[width=0.49\linewidth]{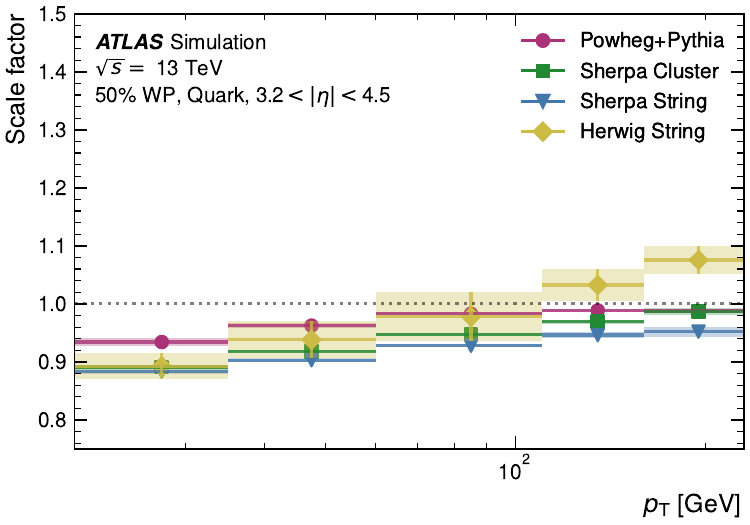}
\label{fig:mc-to-mc_quark_fwd}
}\\
\subfloat[]{
\includegraphics[width=0.49\linewidth]{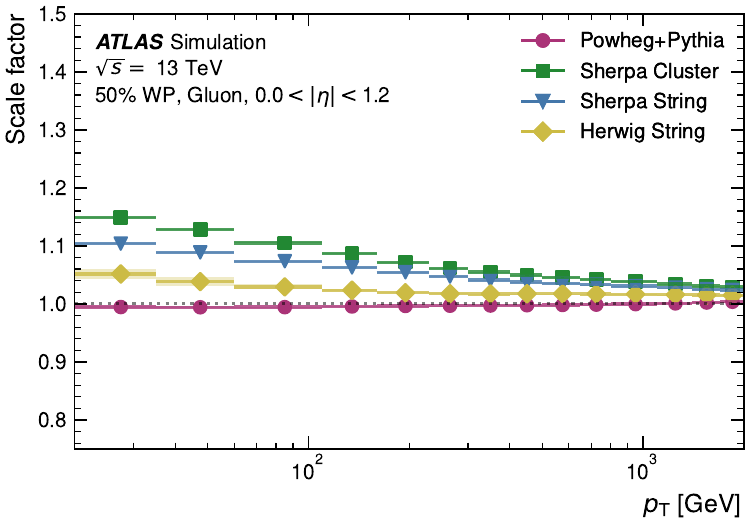}
\label{fig:mc-to-mc_gluon_ctrl}
}\subfloat[]{
\includegraphics[width=0.49\linewidth]{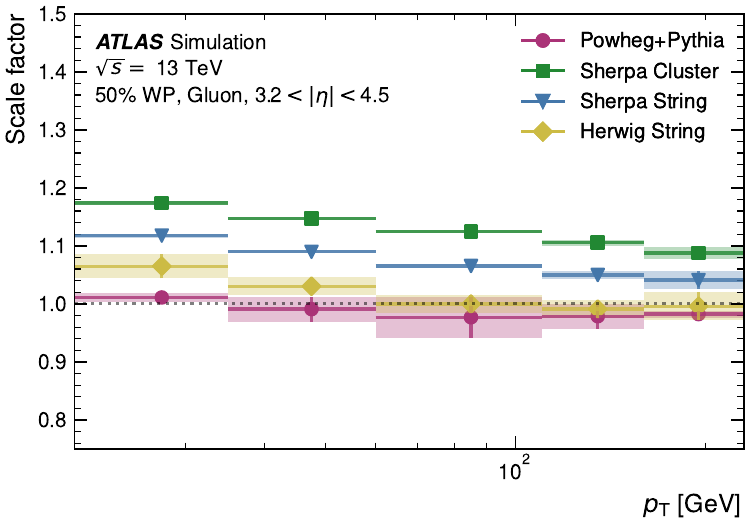}
\label{fig:mc-to-mc_gluon_fwd}
}
\caption{
The MC-to-MC scale factors for
\protect\subref{fig:mc-to-mc_quark_ctrl}~quark jets with $0.0 < |\eta| <  1.2$,
\protect\subref{fig:mc-to-mc_quark_fwd}~quark jets with $3.2 < |\eta| < 4.5$,
\protect\subref{fig:mc-to-mc_gluon_ctrl}~gluon jets with $0.0 < |\eta| <  1.2$, and
\protect\subref{fig:mc-to-mc_gluon_fwd}~gluon jets with $3.2 < |\eta| < 4.5$.
The 50\% WP corresponds to a score threshold defined by a fixed 0.5 quark efficiency in the Pythia sample in each \pt and \abseta bin.
The statistical uncertainties are shown as error bars.
}
\label{fig:mc-to-mc}
\end{figure}


%
\FloatBarrier
\section{Conclusions}
\label{sec:conclusion}
%

This paper presents DeParT, a novel ATLAS $q/g$ tagger leveraging modern deep-learning techniques and directly processing variables derived from jet constituents.
Compared to previous similar ATLAS taggers, the DeParT $q/g$ tagger covers an extended phase space where jets have transverse momentum $\pt > 20~\gev$ and pseudorapidity $|\eta| < 4.5$.

The tagger's performance was evaluated by selecting dijet events from \pp collisions at $\sqrt{s} = 13~\tev$ and $13.6~\tev$ recorded by the ATLAS detector during Runs~2 and 3 of the LHC.
Two methods were employed to extract quark and gluon jet tagger score distributions from data: a novel method called jet topics, and the matrix method used in the previous ATLAS studies.
The two methods yield compatible results, and consistency is observed between the Run-2 and Run-3 data-taking periods.

The efficiencies of the DeParT $q/g$ tagger measured in data, for a tagger working point corresponding to a 50\% quark tagging efficiency in the nominal simulation, were found to range from 0.43 to 0.53 for the quark-initiated jet sample, and from 0.04 to 0.3 for the gluon-initiated one.
These efficiency measurements were compared with the values predicted by Monte Carlo simulations, thereby deriving factors which account for the observed jet-tagging efficiency differences between data and simulation.
The correction factors on the tagging efficiency for quark-initiated jets and on one minus the tagging efficiency for gluon-initiated jets were found to range from 0.88 to 1.30 and 0.61 to 1.05, with uncertainties of 10\%--70\% and 10\%--95\%, respectively.
Tagging efficiency correction factors were also evaluated to account for differences in jet fragmentation modelling, with values ranging from 0.9 to 1.2.

The DeParT $q/g$ tagger represents a significant advancement in the ATLAS $q/g$ tagging capabilities, providing a robust tool for future analyses involving quark- and gluon-initiated jets.
The jet topics method demonstrated its potential as a reliable technique for extracting quark and gluon jet samples from data, offering a promising alternative to traditional methods such as the matrix method.
Systematic uncertainties associated with the jet topics method are smaller than those of the matrix method in an important part of the phase space.
This reduction in systematic uncertainty, particularly pronounced in regions where traditional methods face limitations, establish the jet topics approach as a powerful complementary technique that enhances the robustness of $q/g$ tagging measurements across the full ATLAS detector acceptance.


%
%
%
%
%
\section*{Acknowledgements}
%

%

%
%

%
%

We thank CERN for the very successful operation of the LHC and its injectors, as well as the support staff at
CERN and at our institutions worldwide without whom ATLAS could not be operated efficiently.

The crucial computing support from all WLCG partners is acknowledged gratefully, in particular from CERN, the ATLAS Tier-1 facilities at TRIUMF/SFU (Canada), NDGF (Denmark, Norway, Sweden), CC-IN2P3 (France), KIT/GridKA (Germany), INFN-CNAF (Italy), NL-T1 (Netherlands), PIC (Spain), RAL (UK) and BNL (USA), the Tier-2 facilities worldwide and large non-WLCG resource providers. Major contributors of computing resources are listed in Ref.~\cite{ATL-SOFT-PUB-2025-001}.

We gratefully acknowledge the support of ANPCyT, Argentina; YerPhI, Armenia; ARC, Australia; BMWFW and FWF, Austria; ANAS, Azerbaijan; CNPq and FAPESP, Brazil; NSERC, NRC and CFI, Canada; CERN; ANID, Chile; CAS, MOST and NSFC, China; Minciencias, Colombia; MEYS CR, Czech Republic; DNRF and DNSRC, Denmark; IN2P3-CNRS and CEA-DRF/IRFU, France; SRNSFG, Georgia; BMFTR, HGF and MPG, Germany; GSRI, Greece; RGC and Hong Kong SAR, China; ICHEP and Academy of Sciences and Humanities, Israel; INFN, Italy; MEXT and JSPS, Japan; CNRST, Morocco; NWO, Netherlands; RCN, Norway; MNiSW, Poland; FCT, Portugal; MNE/IFA, Romania; MSTDI, Serbia; MSSR, Slovakia; ARIS and MVZI, Slovenia; DSI/NRF, South Africa; MICIU/AEI, Spain; SRC and Wallenberg Foundation, Sweden; SERI, SNSF and Cantons of Bern and Geneva, Switzerland; NSTC, Taipei; TENMAK, T\"urkiye; STFC/UKRI, United Kingdom; DOE and NSF, United States of America.

Individual groups and members have received support from BCKDF, CANARIE, CRC and DRAC, Canada; CERN-CZ, FORTE and PRIMUS, Czech Republic; COST, ERC, ERDF, Horizon 2020 and Marie Sk{\l}odowska-Curie Actions, European Union; Investissements d'Avenir Labex, Investissements d'Avenir Idex and ANR, France; DFG and AvH Foundation, Germany; Herakleitos, Thales and Aristeia programmes co-financed by EU-ESF and the Greek NSRF, Greece; BSF-NSF and MINERVA, Israel; NCN and NAWA, Poland; La Caixa Banking Foundation, CERCA Programme Generalitat de Catalunya and PROMETEO and GenT Programmes Generalitat Valenciana, Spain; G\"{o}ran Gustafssons Stiftelse, Sweden; The Royal Society and Leverhulme Trust, United Kingdom.

In addition, individual members wish to acknowledge support from Chile: Agencia Nacional de Investigaci\'on y Desarrollo (FONDECYT 1230812, FONDECYT 1240864, Fondecyt 3240661, Fondecyt Regular 1240721); China: Chinese Ministry of Science and Technology (MOST-2023YFA1605700, MOST-2023YFA1609300), National Natural Science Foundation of China (NSFC - 12175119, NSFC 12275265); Czech Republic: Czech Science Foundation (GACR - 24-11373S), Ministry of Education Youth and Sports (ERC-CZ-LL2327, FORTE CZ.02.01.01/00/22\_008/0004632), PRIMUS Research Programme (PRIMUS/21/SCI/017); EU: H2020 European Research Council (ERC - 101002463); European Union: European Research Council (BARD No. 101116429, ERC - 948254, ERC 101089007), European Regional Development Fund (SMASH COFUND 101081355, SLO ERDF), European Union, Future Artificial Intelligence Research (FAIR-NextGenerationEU PE00000013); France: Agence Nationale de la Recherche (ANR-21-CE31-0022, ANR-22-EDIR-0002, ANR-24-CE31-0504-01); Germany: Deutsche Forschungsgemeinschaft (DFG - 469666862, DFG - CR 312/5-2); China: Research Grants Council (GRF); Italy: Ministero dell'Università e della Ricerca (NextGenEU 153D23001490006 M4C2.1.1, NextGenEU I53D23000820006 M4C2.1.1, NextGenEU I53D23001490006 M4C2.1.1, SOE2024\_0000023); Japan: Japan Society for the Promotion of Science (JSPS KAKENHI JP22H01227, JSPS KAKENHI JP22H04944, JSPS KAKENHI JP22KK0227, JSPS KAKENHI JP24K23939, JSPS KAKENHI JP24KK0251, JSPS KAKENHI JP25H00650, JSPS KAKENHI JP25H01291, JSPS KAKENHI JP25K01023); Norway: Research Council of Norway (RCN-314472); Poland: Ministry of Science and Higher Education (IDUB AGH, POB8, D4 no 9722), Polish National Science Centre (NCN 2021/42/E/ST2/00350, NCN OPUS 2023/51/B/ST2/02507, NCN UMO-2019/34/E/ST2/00393, UMO-2022/47/O/ST2/00148, UMO-2023/49/B/ST2/04085, UMO-2023/51/B/ST2/00920, UMO-2024/53/N/ST2/00869); Portugal: Foundation for Science and Technology (FCT); Spain: Generalitat Valenciana (ASFAE/2022/008), Ministry of Science and Innovation (RYC2019-028510-I, RYC2020-030254-I, RYC2021-031273-I, RYC2022-038164-I), Ministerio de Ciencia, Innovación y Universidades/Agencia Estatal de Investigaci\'on (PID2022-142604OB-C22); Sweden: Carl Trygger Foundation (Carl Trygger Foundation CTS 22:2312), Swedish Research Council (Swedish Research Council 2023-04654, VR 2021-03651, VR 2022-03845, VR 2022-04683, VR 2023-03403, VR 2024-05451), Knut and Alice Wallenberg Foundation (KAW 2018.0458, KAW 2022.0358, KAW 2023.0366); Switzerland: Swiss National Science Foundation (SNSF - PCEFP2\_194658); United Kingdom: The Binks Trust, Royal Society (NIF-R1-231091); United States of America: U.S. Department of Energy (ECA DE-AC02-76SF00515), John Templeton Foundation (John Templeton Foundation 63206), Neubauer Family Foundation.

%
%


%
\clearpage
\printbibliography
\clearpage
\input{atlas_authlist}
%

%
%

%

%
%
%
%

\end{document}

%% file: atlas_authlist.tex
 
\begin{flushleft}
\hypersetup{urlcolor=black}
{\Large The ATLAS Collaboration}

\bigskip

\AtlasOrcid[0000-0002-6665-4934]{G.~Aad}$^\textrm{\scriptsize 103}$,
\AtlasOrcid[0000-0001-7616-1554]{E.~Aakvaag}$^\textrm{\scriptsize 17}$,
\AtlasOrcid[0000-0002-5888-2734]{B.~Abbott}$^\textrm{\scriptsize 122}$,
\AtlasOrcid[0000-0002-0287-5869]{S.~Abdelhameed}$^\textrm{\scriptsize 118a}$,
\AtlasOrcid[0000-0002-1002-1652]{K.~Abeling}$^\textrm{\scriptsize 55}$,
\AtlasOrcid[0000-0001-5763-2760]{N.J.~Abicht}$^\textrm{\scriptsize 49}$,
\AtlasOrcid[0000-0002-8496-9294]{S.H.~Abidi}$^\textrm{\scriptsize 30}$,
\AtlasOrcid[0009-0003-6578-220X]{M.~Aboelela}$^\textrm{\scriptsize 45}$,
\AtlasOrcid[0000-0002-9987-2292]{A.~Aboulhorma}$^\textrm{\scriptsize 36e}$,
\AtlasOrcid[0000-0001-5329-6640]{H.~Abramowicz}$^\textrm{\scriptsize 156}$
% ...
\bigskip
\\

% ...
$^{17}$Department for Physics and Technology, University of Bergen, Bergen; Norway.\\
% ...
$^{30}$Physics Department, Brookhaven National Laboratory, Upton NY; United States of America.\\
% ...
$^{36}$$^{(a)}$Facult\'e des Sciences Ain Chock, Universit\'e Hassan II de Casablanca;$^{(b)}$Facult\'{e} des Sciences, Universit\'{e} Ibn-Tofail, K\'{e}nitra;$^{(c)}$Facult\'e des Sciences Semlalia, Universit\'e Cadi Ayyad, LPHEA-Marrakech;$^{(d)}$LPMR, Facult\'e des Sciences, Universit\'e Mohamed Premier, Oujda;$^{(e)}$Facult\'e des sciences, Universit\'e Mohammed V, Rabat;$^{(f)}$Institute of Applied Physics, Mohammed VI Polytechnic University, Ben Guerir; Morocco.\\
% ...
$^{45}$Physics Department, Southern Methodist University, Dallas TX; United States of America.\\
% ...
$^{49}$Fakult\"{a}t Physik, Technische Universit{\"a}t Dortmund, Dortmund; Germany.\\
% ...
$^{55}$II. Physikalisches Institut, Georg-August-Universit\"{a}t G\"ottingen, G\"ottingen; Germany.\\
% ...
$^{103}$CPPM, Aix-Marseille Universit\'e, CNRS/IN2P3, Marseille; France.\\
% ...
$^{118}$$^{(a)}$New York University Abu Dhabi, Abu Dhabi;$^{(b)}$United Arab Emirates University, Al Ain; United Arab Emirates.\\
% ...
$^{122}$Homer L. Dodge Department of Physics and Astronomy, University of Oklahoma, Norman OK; United States of America.\\
% ...
$^{156}$Raymond and Beverly Sackler School of Physics and Astronomy, Tel Aviv University, Tel Aviv; Israel.\\
% ...
$^{a}$ Also at Affiliated with an institute formerly covered by a cooperation agreement with CERN.\\
$^{b}$ Also at An-Najah National University, Nablus; Palestine.\\
$^{c}$ Also at Borough of Manhattan Community College, City University of New York, New York NY; United States of America.\\
$^{d}$ Also at Center for Interdisciplinary Research and Innovation (CIRI-AUTH), Thessaloniki; Greece.\\
$^{e}$ Also at Centre of Physics of the Universities of Minho and Porto (CF-UM-UP); Portugal.\\
$^{f}$ Also at CERN, Geneva; Switzerland.\\
% ...

\end{flushleft}
